\newif\ifsingle

\ifsingle
\documentclass[11pt,draftclsnofoot]{IEEEtran}
\else		
\documentclass[10pt,final, twocolumn]{IEEEtran}
\fi

\newif\ifFullVersion
\FullVersiontrue 

\usepackage{amsmath,graphicx}
\usepackage{amssymb}
\usepackage{acronym}
\usepackage{multirow}
\usepackage[belowskip=-10pt,aboveskip=1.8pt]{caption}
\usepackage{standalone}
\usepackage{subcaption}
\usepackage{tikz}
\usepackage{url,cite}
\usepackage{verbatim}
\usepackage[bookmarks,colorlinks]{hyperref}
\usepackage{soul, xcolor}
\usepackage{cancel}
\usepackage{enumitem}
\usepackage[ruled,linesnumbered,vlined]{algorithm2e}
\SetKwInput{KwData}{\textbf{Init}} 

\newcommand{\myVec}[1]{{\boldsymbol{#1}}}
\newcommand{\myMat}[1]{{\boldsymbol{#1}}}
\newcommand{\mySet}[1]{\mathcal{#1}}

\newtheorem{theorem}{Theorem}

\acrodef{nn}[NN]{Neural Network}
\acrodef{pga}[PGA]{projected gradient ascent}
\acrodef{snr}[SNR]{signal-to-noise ratio}
\acrodef{bs}[BS]{base station} 
\acrodef{cpu}[CPU]{centralized processing unit} 
\acrodef{mimo}[MIMO]{multiple-input multiple-output}
\acrodef{awgn}[AWGN]{additive white Gaussian noise} 
\acrodef{cpu}[CPU]{central processing unit} 
\acrodef{ml}[ML]{machine learning} 
\acrodef{adc}[ADC]{analog-to-digital converter} 
\acrodef{mse}[MSE]{mean-squared error}
\acrodef{cnn}[CNN]{convolutional neural network}
\acrodef{iot}[IOT]{internet-of-things}
\acrodef{rmse}[RMSE]{root mean squared error}
\acrodef{rmspe}[RMSPE]{root mean squared periodic error}
\acrodef{mmse}[MMSE]{{minimum mean-squared error}}
\acrodef{lmmse}[LMMSE]{{linear} MMSE}
\acrodef{mle}[MLE]{maximum likelihood estimation}
\acrodef{snr}[SNR]{signal-to-noise ratio} 
\acrodef{csi}[CSI]{channel state information}
\acrodef{quadriga}[QuaDRiGa]{Quasi-Deterministic Radio channel Generator}

\usetikzlibrary{trees,arrows}

\setstcolor{blue}

\definecolor{NewColor}{rgb}{0,0,0}
 
\title{Rapid and Power-Aware Learned Optimization for Modular Receive Beamforming}

\author{
\IEEEauthorblockN{Ohad Levy and Nir Shlezinger
\thanks{
Parts of this work were presented in the 2024 IEEE International Conference on Acoustics, Speech, and Signal Processing (ICASSP) as the paper~\cite{levy2024rapid}.
This work was supported by the Israeli Innovation Authority. 
The authors are with the School of ECE, Ben-Gurion University of the Negev, Israel (e-mail: levyoha@post.bgu.ac.il; nirshl@bgu.ac.il). 
}}
}

\begin{document}

\maketitle
%
%
\begin{abstract} 
\Ac{mimo} systems play a key role in wireless communication technologies. 
A widely considered approach to realize scalable \ac{mimo} systems involves architectures comprised of multiple separate modules, each with its own beamforming capability. Such models accommodate cell-free massive \ac{mimo} and partially connected hybrid \ac{mimo} architectures. A core issue with the implementation of modular \ac{mimo} arises from the need to rapidly set the beampatterns of the modules, while maintaining their power efficiency. This leads to challenging constrained optimization that should be repeatedly solved on each coherence duration. 
In this work, we propose a power-oriented optimization algorithm for beamforming in uplink modular hybrid \ac{mimo} systems, which learns from data to operate rapidly. We derive our learned optimizer by tackling the rate maximization objective using projected gradient ascent steps with momentum. We then leverage data to tune the hyperparameters of the optimizer, allowing it to operate reliably in a fixed and small number of iterations while completely preserving its interpretable operation. We show how power efficient beamforming can be encouraged by the learned optimizer, via boosting architectures with low-resolution phase shifts and with deactivated analog components.  Numerical results show that our learn-to-optimize method notably reduces the number of iterations and computation latency required to reliably tune modular \ac{mimo} receivers, and that it allows obtaining desirable balances between power efficient designs and throughput.
\end{abstract}

\acresetall

\section{Introduction}
\label{sec:intro}
Massive \ac{mimo} technology is at the core of future wireless communications, with extremely large implementations expected to enable meeting the constantly growing demands in throughput and coverage~\cite{wang2024tutorial,bjornson2024towards}. However, implementing communications with hundreds or thousands of antenna elements gives rise to several challenges, including issues related to cost, power consumption, deployment, and scalability~\cite{bjornson2019massive}.
\textcolor{NewColor}{To overcome these challenges, a leading approach to massive \ac{mimo}  adopts {\em modular designs}. In such cases, a \ac{mimo}  array is implemented using many distinct modules, each equipped with a (smaller) antenna array, and possibly possesses local beamforming capabilities.
These  architectures encompass various forms of emerging \ac{mimo} technologies, including extremely large arrays~\cite{li2022near,li2024multi}, modular low-frequency deployments~\cite{jeon2021mimo} distributed cell-free massive \ac{mimo}~\cite{bjornson2020scalable}, centralized interconnections of multiple panels~\cite{7931558,huang2018multi}, and even radio stripes~\cite{lopez2022massive}.}

While modular implementations of massive \ac{mimo} alleviate cost and power issues and support distributed implementation, they also introduce design and algorithmic challenges. In particular, beamforming is not carried out by a single \ac{cpu} that manipulates the signal at each element; instead, it is divided between the \ac{cpu} and a set of separate arrays termed {\em panels}, that are possibly constrained in their hardware capabilities, and are likely to be subject to strict power constraints. Moreover, the setting of the modular beampattern is done on each coherence duration, necessitating rapid adjustments of the modules~\cite{raviv2024adaptive}.

Since modular beamforming is divided between a \ac{cpu} and multiple panels, it can be viewed as a form of {\em hybrid beamforming}~\cite{molisch2017hybrid}. Hybrid beamforming divides the processing of the signals between conventional digital processing and an additional constraint antenna interface, typically representing analog hardware~\cite{mendez2016hybrid}. Such hybrid \ac{mimo} is often studied in the context of RF chain reduction~\cite{ioushua2019family, gong2019rf}, most commonly assuming phase shifter based analog processing~\cite{park2017dynamic}. However, hybrid \ac{mimo} can  also represent emerging antenna architectures based on true-time-delay~\cite{li2022rainbow}, vector modulators~\cite{tasci2022robust}, metasurfaces~\cite{shlezinger2021dynamic}, or leaky waveguides~\cite{gabay2024leaky}. 

The main task of hybrid beamforming can be viewed as the conversion of \ac{csi} into a beampattern achievable through a divisible architecture~\cite{elbir2022twenty}. Various methods have been proposed for tackling this task (see, e.g., overview in~\cite{shlezinger2023ai}). As hybrid beamforming can be represented as a constrained optimization problem, a conventional approach achieves this via iterative optimizers, see, e.g.,~\cite{yu2016alternating, sohrabi2016hybrid,qiao2020alternating}. However, these approaches typically involve lengthy optimization with numerous iterations, which can be challenging to carry out efficiently within the extremely short time frames required for determining beamforming (e.g., on the order of $0.1$ milliseconds~\cite{shlezinger2023ai}).

To facilitate rapid beamforming setting, recent works proposed using deep learning tools, taking advantage of their ability to learn to tackle challenging optimization problems with fixed and limited latency~\cite{zappone2019wireless,chen2021learning}. This can be achieved by training conventional deep learning architectures, such as \acp{cnn}, to map \ac{mimo} \ac{csi} into  analog and digital precoders~\cite{elbir2019hybrid,dong2020framework}, or by
combining optimization with deep learning for hybrid beamformers~\cite{lavi2023learn,nguyen2023deep,balevi2021unfolded,shi2022deep} \textcolor{NewColor}{and for fully digital beamformers~\cite{pellaco2021matrix}} as a form of model-based deep learning~\cite{shlezinger2022model}. However, while hybrid beamforming can be viewed as \textcolor{NewColor}{encompassed} by modular beamforming, its associated algorithms are not tailored for exploiting modular structures, and do not account for the limited power considerations imposed in such forms of distributed \ac{mimo} systems. Moreover, current learning-aided designs often focus on a given configuration, and the optimizer needs to be retrained when using a different array (or adding another module), limiting scalability. Finally, existing works typically focus on downlink beamforming, with the uplink being much less studied.

Uplink modular beamforming was considered in \cite{alegria2021trade,alegria2023trade}. There, the focus was to characterize trade-offs of its decentralization under narrowband communications and sparsity-based power-oriented constraints for loss-less processing, rather than establishing algorithms for tuning modular beamformers. In practice, the regimes of interest are typically those where the distributed and power constrained operation is likely to yield some rate loss compared to centralized fully digital \ac{mimo} systems. This gives rise to the need to design methods for tuning hybrid modular receive beamformers for such practical and challenging settings.

In this work, we develop a rapid and power-aware tuning algorithm for setting modular receive \ac{mimo} beamformers. 
Our approach combines principled optimization with machine learning  via model-based deep learning~\cite{shlezinger2020model} to realize  uplink modular beamforming optimizers operating reliably with fixed and low latency. By doing so, we are able to utilize data to optimize the optimizer while maintaining its comprehensibility and transferability across different configurations. 

We focus on modular beamformers implemented using conventional phase shifters, and leverage the interpretable operation of the optimizer to $(i)$ facilitate scalable design, where the same learned optimizer can be utilized with different numbers of modules; and $(ii)$  encourage power efficient designs for each panel. The latter is achieved by enabling deactivation and low-resolution implementation of the active phase shifters at each module. 
We formulate the modular uplink beamforming task as a constrained maximization of the sum-rate objective in multiple-access channels. We then specialize \ac{pga} steps with momentum for the considered setting, being a suitable candidate accelerated iterative optimizer for directly optimizing the rate objective, based on which we develop our learning-aided optimizer. 

Our main contributions are summarized as follows:
\begin{itemize}
    \item \textbf{First-order optimization for modular receive beamforming:}
We formulate the receive modular beamforming task as a constrained optimization, and derive the corresponding first-order iterative optimizer based on \ac{pga} steps. 
\textcolor{NewColor}{The resulting optimization formulation differs from previous works that focused on the downlink, namely, transmit beamforming.}
We specifically characterize the corresponding gradient steps and projection methods, which enable (lengthy) tuning of modular receive beamformers.
    \item \textbf{Rapid and scalable learned optimization:}  
To enable rapid tuning, we convert the identified \ac{pga} steps into a machine learning model operating with a fixed and low latency. This is achieved by fixing the number of iterations while using the hyperparameters of \ac{pga} with momentum as trainable machine learning parameters. We show that the interpretable operation of the resulting design allows it to scale to different numbers of modules (potentially much larger than those used during training), thus facilitating scalability, \textcolor{NewColor}{and that its training procedure can be adapted to enhance robustness to noisy channel estimates.}
    \item \textbf{Power-aware learned optimization:}  
We leverage the interpretable structure of the resulting machine learning model, comprised of \ac{pga} iterations, to incorporate two forms of power-aware considerations: the ability to boost sparse modular architectures with deactivated phase shifters, and the support of low-resolution phase shifters. As the resulting constraints render \ac{pga} non-differentiable, we approximate its operation for learning purposes via surrogate projection mappings that enable its representation as a machine learning architecture.
    \item \textbf{Extensive experimentation:}
We extensively evaluate our learned optimizer on both synthetic channels and channels obtained from the physically compliant \ac{quadriga} model~\cite{6758357}. 
\textcolor{NewColor}{Our numerical studies systematically show that learned optimization enables reliable tuning of modular beamformers with reductions of $7-15 \times$ in iterations compared to conventional optimizers. Moreover, we show that the power-aware design enables efficient operation, typically with only a minor degradation in accuracy compared to unconstrained designs. It is also shown that under some forms of limited designs, our method achieves superior performance in just $10$ iterations, significantly outperforming other methods, which required $500$ iterations to converge to a suboptimal solution.
Additionally, the proposed method exhibits improved robustness under \ac{csi} estimation errors, demonstrating a notably softer degradation in performance compared to traditional approaches as the \ac{csi} noise variance increases. These results reveal the versatility of our approach, which balances computational efficiency and accuracy, even under challenging power, hardware, or imperfect \ac{csi} conditions.}
\end{itemize}

The rest of this paper is organized as follows: Section~\ref{sec:System Model} presents the system model of uplink modular beamforming and formulates its design problem. Section~\ref{sec:Unfolded Beamforming} presents our learned optimization method without accounting for power considerations, which are introduced into the method in Section~\ref{sec:Power Aware Beamforming}. Empirical results are reported in Section~\ref{sec:Experimental Study}, and Section \ref{ssec:conclusions}~concludes the paper.

Throughout this paper, we use boldface lowercase and boldface uppercase letters for column vectors and matrices, respectively, while $[\myMat{M}]_{i,j}$ is the $(i,j)$th element of a matrix $\myMat{M}$.
We use calligraphic fonts for sets, while $\mathbb{Z}$ and $\mathbb{C}$ denote the integer and complex numbers, respectively. The operations $(\cdot)^T$ and $(\cdot)^H$ are used for transpose and conjugate transpose, respectively. $\myMat{I}_N$ refers to the identity matrix of size $N$. The gradient of the function $f(\cdot)$ with respect to the matrix $\myMat{W}$ is denoted by $\nabla_{\myMat{W}}f(\cdot)$, while $\delta_{\cdot,\cdot}$ is the Kronecker delta function, and $\measuredangle (\cdot)$ is the (element-wise) phase operator (in radians).
%
%
\section{System Model}
\label{sec:System Model}

In this section we introduce the system model for receive beamforming with modular hybrid \ac{mimo} receivers. We commence with presenting the considered models for the communication channel and the receiver processing in Subsections~\ref{ssec:Communicaiton}-\ref{ssec:receivers}, respectively, after which we formulate the beamforming design problem in Subsection~\ref{ssec:problem}.

\subsection{Communication System}
\label{ssec:Communicaiton}
We consider a multi-band multi-user uplink \ac{mimo} system. The setting involves $K$ single antenna users that communicate with a (possibly distributed) \ac{mimo} receiver,  over $B$ frequency bins. The transmitted signals are received by $M$ antenna elements, that are divided into multiple panels. The channel output at the $b$th frequency
is given by
\begin{equation}
    \label{eqn:RxSignal}
    \myVec{y}[b]=\myMat{H}[b] \myVec{s}[b] + \myVec{w}[b], \qquad b=1,\ldots,B.
\end{equation}
In \eqref{eqn:RxSignal}, $\myMat{H}[b]\in \mathbb{C}^{M\times K}$ is the channel matrix at frequency bin $b$, while  $\myVec{s}[b] \in \mathbb{C}^K$ represents the transmitted symbols, which are comprised of i.i.d. entries with power $\rho_s>0$. The signal $\myVec{w}[b] \in \mathbb{C}^M$ is \ac{awgn} with variance $\sigma_w^2>0$.

\subsection{Modular Hybrid Receive Processing}
\label{ssec:receivers}
\begin{figure}
    \centering
    \includegraphics[width=\columnwidth]{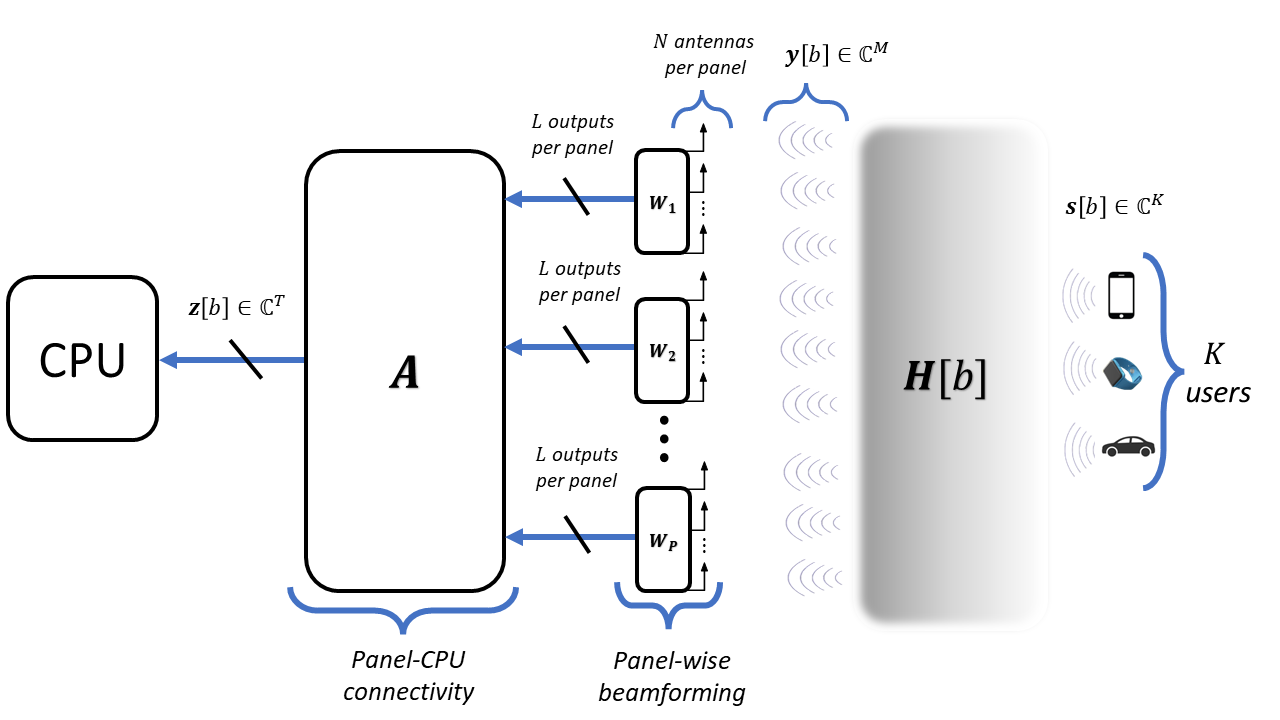}
    \caption{Modular hybrid receive processing chain.}
    \label{fig:decentralized}
\end{figure}

The signal received by the $M$ antenna elements is processed by a centralized \ac{cpu}. The resulting digitally processed signal is obtained from the channel output in \eqref{eqn:RxSignal} by undergoing {\em modular receive beamforming}, where the $M$ antennas are divided into $P$ panels. This processing, illustrated in Fig.~\ref{fig:decentralized}, includes two stages: $(i)$ Panel-wise beamforming, and $(ii)$ Panel-\ac{cpu} connectivity.

\subsubsection{Panel-Wise Beamforming}
The first stage is composed of receive beamforming performed by each panel separately. Each panel has $N$ antennas and $L$ output ports, where the data processing takes place within the $p$th antenna panel and is described via a ${N\times L}$ frequency-invariant matrix $\myMat{W}_p$, for each $p=1,\ldots,P$. The panels are assumed to be hardware limited, which we model using the set of feasible mappings $ \mySet{W} \subset \mathbb{C}^{N\times L}$, such that $\myMat{W}_p \in \mySet{W}$. 

We consider the following types of analog constraints:
\begin{enumerate}[label={C\arabic*}]
	\item \label{itm:unconstratined} {\em Unconstrained Phase Shifters}: Here, the analog processing is comprised of phase shifters, where the phase of component can be set arbitrarily, i.e., 
	\begin{equation}
	\label{eqn:PhasShiftConst}
	\mySet{W}_{\rm uph} = \left\{\myMat{W}\in \mathbb{C}^{N \times L}: \left|[\myMat{W}]_{n,l}\right| = 1, \forall (n,l)\right\}.
	\end{equation}
	\item \label{itm:sparse}  {\em Sparsified Phase Shifters}: As phase shifters are active components with non-negligible power consumption, a candidate approach to maintain power-efficient operation at the panels is to allow some of the phase shifters to be turned off~\cite{tasci2022robust}. This type of feasible set is represented as
	\begin{equation}
	\label{eqn:PhasShiftBinary}
 \hspace{-0.2cm}
	\mySet{W}_{\rm sph} = \left\{\myMat{W}\in \mathbb{C}^{N \times L}: \left|[\myMat{W}]_{n,l}\right| \in \{0,1\}, \forall (n,l) \right\}.
	\end{equation}
	While the formulation in \eqref{eqn:PhasShiftBinary} does not indicate how many phase shifters are inactive, such a requirement can be imposed by constraining the $\ell_0$ norm of $\myMat{W}$ to achieve a desired amount of inactive components. 
	
	
	%
	\item \label{itm:quant}  {\em Quantized Phase Shifters}: An alternative  approach to limit the power consumption on the panel side is to restrict the phase shifters to support only a discrete set of feasible phases, i.e., support quantized phases shifts~\cite{deng2019mmwave}. Such a constraint with $Q$-level quantized phases is modeled by the feasible set
	\begin{align}
	\label{eqn:PhasShiftQuant}
	\mySet{W}_{\rm qps}^{Q}  = \big \{ \myMat{W} \in \mathbb{C}^{N \times L} :    &\frac{\measuredangle[\myMat{W}]_{n,l}}{2\pi} \cdot Q \in \mathbb{Z} ,
	\notag \\
	& \left|[\myMat{W}]_{n,l}\right| = 1,  \forall (n,l) \big \}.
	\end{align}
\end{enumerate}
Note that for all considered feasible sets, the phase shifting components remain frequency-agnostic, i.e., $\myMat{W}_p$ is invariant of the frequency index $b$. \textcolor{NewColor}{It is also noted that the constrained form of receive beamforming in general cannot realize conventional simple receive beamformers, such as the zero-forcing of minimal mean squared-error beamformers~\cite[Ch. 8.3]{tse2005fundamentals}.}
Moreover, we note that  approaches \ref{itm:sparse} and \ref{itm:quant}, which aim to improve the power efficiency of the panels, are complementary. Accordingly, one can restrict the analog processing at the panels side to be both sparse and quantized, i.e., have some of the phase shifters deactivated, while the active ones take discrete phase values.

\subsubsection{Panel-CPU Connectivity}
We model the connectivity to the \ac{cpu}  via the binary matrix $\myMat{A}\in \{0,1\}^{L\cdot P \times T}$, with $T$ being the number of \ac{cpu} inputs. This matrix is also frequency invariant, and its entry $[\myMat{A}]_{l \cdot p,t}$ denotes whether the $l$th output of the $p$th panel is connected to the $t$th \ac{cpu} input. 
\textcolor{NewColor}{The matrix $\myMat{A}$ represents  the  physical connections (e.g., wiring) between the panels and the centralized processor, and is thus assumed to fixed by the modular \ac{mimo} infrastructure.}
%
The \ac{cpu} input  $\myVec{z}[b]$ is related to the channel output $\myVec{y}[b]$ via
\begin{equation}
\label{eqn:HBF}
    \myVec{z}[b]=\myMat{A}^H \myMat{W}^H  \myVec{y}[b],
\end{equation}
where $ \myMat{W}= {\rm blkdiag}\{\myMat{W}_1,\ldots,\myMat{W}_P\}$. This constraint comes from the decentralized structure of our communication system, which takes place within each panel. 

\textcolor{NewColor}{The above model for modular hybrid receive processing accommodates both distributed \ac{mimo} settings with separated panels~\cite{bjornson2020scalable}, as well as massive \ac{mimo} architectures where the panels are collocated on the same platform~\cite{huang2018multi}. The distinction between such settings lies in the resulting channel matrices $\{\myMat{H}[b]\}$, which we do not constrain to take a specific form. Moreover, our model can be viewed as specializing different forms of hybrid \ac{mimo} systems when $\myMat{A}$ is the identity matrix, e.g., having $L=1$ in such cases specializes the common form of partially-connected hybrid beamformers, while setting $P=1$ realizes fully-connected hybrid beamformers~\cite{ioushua2019family}.}

\subsection{Problem Formulation}
\label{ssec:problem}
We aim to design the modular receive beamformer detailed in \eqref{eqn:HBF} to maximize the achievable rate. To formulate this, we define the equivalent channel at the $b$th frequency bin as 
    $\myVec{G} \triangleq \myMat{W} \myMat{A} \in \mathbb{C}^{M\times T}$.
For the uplink setup detailed in Subsection~\ref{ssec:Communicaiton}, the average achievable sum-rate is given by (see, e.g.,~\cite[Thm. 1]{shlezinger2019dynamic})
\begin{align}
    R\left(\myMat{G}; \{\myMat{H}[b]\}\right) = \frac{1}{B}\sum_{b=1}^B \log  \bigg|\myMat{I}_M +&\frac{\rho_s}{N_0}\myMat{G}\big(\myMat{G}^H\myMat{G}\big)^{-1}\notag\\
    &\myMat{G}^H
    \myMat{H}[b]\myMat{H}^H[b]\bigg|.
    \label{eqn:Rate}
\end{align}

Using the rate expression in \eqref{eqn:Rate}, we can formulate the setup of modular receive beamforming as a constrained optimization problem.
Since the matrix $\myMat{A}$ represents the hardware connectivity between the panels, it is assumed to be constant, thus not optimized.
Accordingly, for a given channel realization $\{\myMat{H}[b]\}_{b=1}^B\}$, the modular beamforming task is formulated as the recovery of 
\begin{align}
   \label{maximization_R}
    \myMat{W}^\star &= \mathop{\arg \max}\limits_{\myMat{W}= {\rm blkdiag}\{\myMat{W}_1,\ldots,\myMat{W}_P\}} R(\myMat{W}\myMat{A};  \{\myMat{H}[b]\}) \\
    &\text{subject to } \myMat{W}_p \in \mySet{W}, \quad \forall p\in\{1,\ldots,P\}. \notag
\end{align}

The problem as it is formulated in \eqref{maximization_R} considers a generic setting for the feasible panel-wise  computation set $\mySet{W}$. As such, it accommodates both the unstructured setting \ref{itm:unconstratined} as well as the power-oriented \ref{itm:sparse}-\ref{itm:quant}. However, when required to operate with sparse panels \ref{itm:sparse}, one is likely to also impose an additional constraint on the overall number of active phase shifters  in $\myMat{W}$. Similarly, imposing \ref{itm:quant} assumes that one operates with a fixed and known resolution dictated by $Q$.

In some cases (e.g., when $T \geq \max (\lfloor M\frac{K-L}{K}+1 \rfloor,K)$, $B=1$, and while allowing $\mySet{W} = \mathbb{C}^{N \times L}$), one can identify the optimal setting~\cite{alegria2021trade}. Even in such cases,  tackling \eqref{maximization_R} typically involves an iterative optimization procedures. In practice, the modular beamformer must be set within a coherence duration (which can be of the order of $0.1$ milliseconds~\cite{shlezinger2023ai}). To account for this requirement, we limit the number of iterations for these calculations, allowing at most $J$ iterations. To facilitate identifying an optimizer that can reliably tackle \eqref{maximization_R} within $J$ iterations, we assume access to a data set $\mySet{D}$ composed of $|\mySet{D}|>0$ past channel realizations, i.e., 
 \begin{equation}
     \label{eqn:DataSet}
     \mySet{D}=\left\{\{\myMat{H}_i[b]\}_{b=1}^{B} \right\}_{i=1}^{|\mySet{D}|}.
 \end{equation}

\section{Unconstrained Unfolded Modular Beamforming}
\label{sec:Unfolded Beamforming}
In this section, we study modular beamforming design while focusing on {\em unconstrained phase shifters}, i.e., constraint~\ref{itm:unconstratined}. The  unfolded modular beamforming algorithm, which is geared towards enabling rapid adaptation, serves as a starting point for considering the more restrictive constraints \ref{itm:sparse}-\ref{itm:quant} in Section~\ref{sec:Power Aware Beamforming}. In particular, we  first formulate our baseline optimizer of \ac{pga} with momentum in Subsection~\ref{ssec:PGA with momentum}. Then, we convert it into a machine learning model in Subsection~\ref{ssec:unfoldedPGA}, and provide a discussion in Subsection~\ref{ssec:discussion}.

\subsection{PGA with Momentum}
\label{ssec:PGA with momentum}
A direct approach to tackle \eqref{maximization_R} employs first-order methods~\cite{boyd2004convex}, e.g., \ac{pga}. This scheme starts from some initial guess $\myMat{W}_0$, and iteratively takes gradient steps towards the maximization of \eqref{eqn:Rate}, projected to meet the constraints. \ac{pga} relies on the ability to compute the gradient of $R(\myMat{W}\myMat{A};  \{\myMat{H}[b]\}) $ with respect to $\myMat{W}$, denoted $\nabla_{\myMat{W}}R(\myMat{W},\myMat{A})$. The gradient of the rate is stated in Theorem~\ref{thm:dR_dW}.
\begin{theorem}
The gradient of \eqref{eqn:Rate} with respect to $\myMat{W}$ is 
    \label{thm:dR_dW}
    \begin{align} 
        \nabla_\myMat{W}& R(\myMat{W},\myMat{A}) =
        \frac{1}{B}\sum_{b=1}^B
        \bigg(\frac{\rho_s}{\sigma_w^2}\myMat{A} \Big((\myMat{G}^H\myMat{G})^{-1}\myMat{G}^H \notag \\
        & \times \myMat{H}[b]\myMat{H}[b]^H\Big)
        \Big(\myMat{I} +\frac{\rho_s}{\sigma_w^2}   \big(\myMat{G}(\myMat{G}^H\myMat{G})^{-1}
        \myMat{G}^H \notag \\
        & \times \myMat{H}[b]\myMat{H}[b]^H\big)\Big)^{-1}
        \Big(\myMat{I}\!-\!\myMat{G}    (\myMat{G}^H\myMat{G})^{-1}\myMat{G}^H\Big)\bigg) ^H.
        \label{eqn:dR_dW}
    \end{align}
\end{theorem}

\begin{IEEEproof}
    The proof is given in Appendix \ref{app:Proof1}. 
\end{IEEEproof}

\smallskip
In addition to the gradient in~\eqref{eqn:dR_dW}, one should also be able to project onto the feasible constraints in \eqref{maximization_R}  to formulate  \ac{pga}.
For the non-power-constrained feasible set \eqref{eqn:PhasShiftConst}, the projection operator for a matrix $\myMat{W}$ based on \ref{itm:unconstratined} is
\begin{align}
[\mySet{P}_{\mySet{W}_{\rm uph}}(\myMat{W})]_{i,j} &= \frac{[\myMat{W}]_{i,j}}{\left|[\myMat{W}]_{i,j}\right|} \cdot \delta_{\left\lfloor\frac{i}{N}\right\rfloor, \left\lfloor\frac{j}{L}\right\rfloor}.
\label{eqn:Projection}
\end{align}
In \eqref{eqn:Projection}, the element-wise operation divides each entry of the matrix $\myMat{W}$ by its own magnitude, and then sets it to zero if it falls out of the main block diagonal. This effectively keeps each panel-wise processing made up of phase-shifters only.

Although first-order methods often require a large number of iterations to converge, they can typically be accelerated by introducing momentum. Momentum is also useful for dealing with non-convex objectives, as in \eqref{maximization_R}, since it can prevent getting trapped in local minima and saddle points. The resulting \ac{pga} with momentum is summarized as Algorithm~\ref{alg:pga with momentum}.

\begin{algorithm}
    \caption{PGA with Momentum}
    \label{alg:pga with momentum} 
    \SetKwInOut{Initialization}{Init}
    \Initialization{Initialize $\myMat{W}_{-1}$ as zeros; randomize $\myMat{W}_{0}$.
    \newline Set iterations $J$, hyperparameters $\{\mu_j,\beta_j\}_{j=0}^{J-1}$;.}
    \SetKwInOut{Input}{Input}  
    \Input{$\{\myMat{H}[b]\}_{b=1}^B$, $\myMat{A}$}  
    {
        \For{$j = 0, 1, \ldots, J-1$}{%
                Compute gradient $\nabla_\myMat{W}R(\myMat{W}_j,\myMat{A})$ via \eqref{eqn:dR_dW}\;
                Take gradient step $\myMat{W}_{j+1}\leftarrow \myMat{W}_j + \mu_j\nabla_{\myMat{W}}R(\myMat{W}_j,\myMat{A})$\;
                Add momentum  $\myMat{W}_{j+1}\leftarrow \myMat{W}_{j+1}+ \beta_j (\myMat{W}_j-\myMat{W}_{j-1})$\;
                Project $\myMat{W}_{j+1} \leftarrow \mySet{P}_{\mySet{W}_{\rm uph}}(\myMat{W}_{j+1})$  via \eqref{eqn:Projection}\;  
                }
        \KwRet{$\myMat{W}_J$}
  }
\end{algorithm}
%

\subsection{Unfolded PGA with Momentum}
\label{ssec:unfoldedPGA}
Algorithm~\ref{alg:pga with momentum} allows tackling \eqref{maximization_R} while meeting the requirement to operate with $J$ iterations. However, iterative optimizers such as \ac{pga} are designed to tackle convex optimization problems and to operate with asymptotically large number of iterations. There, convergence can be guaranteed with fixed  hyperparameters $\{\mu,\beta\}$ (i.e., independent of the iteration index $j$), which can be tuned manually. In Algorithm~\ref{alg:pga with momentum}, we apply this convex optimizer to a non-convex problem \eqref{maximization_R} while operating with a fixed and small number of iterations $J$. In such cases, the  setting of the hyperparameters $\{\mu_j,\beta_j\}$ greatly affects performance~\cite[Ch. 9]{boyd2004convex}. This motivates leveraging the available  data in \eqref{eqn:DataSet} 
to tune these hyperparameters, by converting Algorithm~\ref{alg:pga with momentum} into a discriminative machine learning model~\cite{shlezinger2022discriminative}, and proposing a dedicated learning mechanism. We next detail the resulting machine learning architecture, its proposed training procedure, \textcolor{NewColor}{how its trainability can leveraged to increase robustness to \ac{csi} errors,} and how it can scaled to large modular settings.

\subsubsection{Architecture}
\label{subssec:Architecture}
Algorithm~\ref{alg:pga with momentum} operates with exactly $J$ iterations. It can thus be readily converted into a machine learning model via deep unfolding for learning hyperparameters~\cite{shlezinger2022model}. Here, each iteration is treated as a layer in a multi-layer architecture, whose trainable weights are the hyperparameters. 

\begin{figure}
    \centering
    \includegraphics[width=\columnwidth]{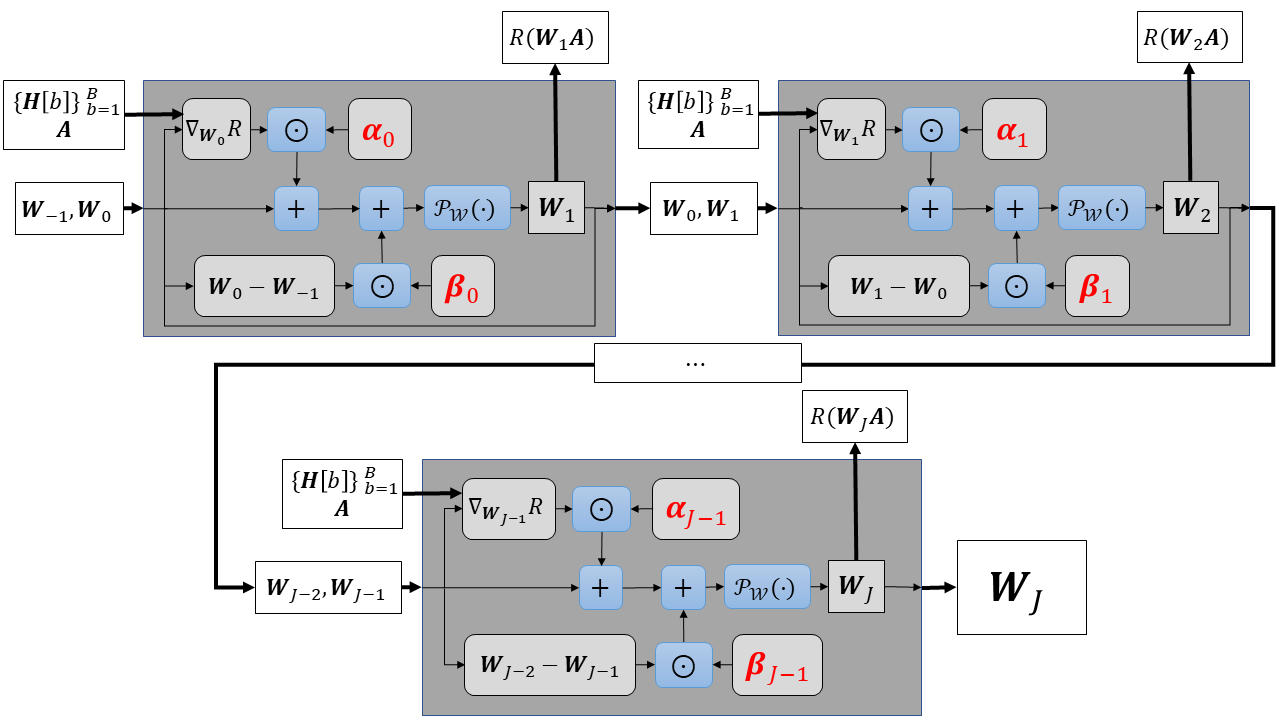}
    \caption{Architecture, trainable parameters  marked with \textcolor{red}{red}.} \label{fig:Deep Unfolding}
\end{figure}

We increase the abstractness of the trainable architecture by allowing it to use different values of $\mu$ and $\beta$ not just per iteration, but also per entry in $\myMat{W}$. Accordingly, the scalar values  $\{\mu_j, \beta_j\}$ that multiply matrix quantities in Algorithm~\ref{alg:pga with momentum} are replaced with block-diagonal matrices $\myMat{\alpha}_j, \myMat{\beta}_j$ that multiply the matrix quantities in Algorithm~\ref{alg:pga with momentum} element-wise.
The resulting architecture obtained from Algorithm~\ref{alg:pga with momentum} is illustrated in Fig.~\ref{fig:Deep Unfolding}, and has $2\cdot J \cdot P \cdot N \cdot L$ trainable parameters,  written as
 \begin{equation}
 \label{eqn:trainableParams}
    \myVec{\theta}=  \{\myMat{\alpha}_j, \myMat{\beta}_j\}_{j=0}^{J-1}.
\end{equation}

\subsubsection{Training}
\label{subssec: Training}
Converting Algorithm~\ref{alg:pga with momentum} into a machine learning model facilitates tuning $\myVec{\theta}$  in \eqref{eqn:trainableParams},
which are the hyperparameters of Algorithm~\ref{alg:pga with momentum}. This is achieved using the dataset $\mySet{D}$ in \eqref{eqn:DataSet}, containing past channel realizations taken from the same communication system. Letting $\myMat{W}_J(\myVec{\theta}; \{\myMat{H}[b]\})$ be the modular beamforming matrix computed by Algorithm~\ref{alg:pga with momentum} with parameters $\myVec{\theta}$ \textcolor{NewColor}{when applied to channel $\{\myMat{H}[b]\}$}, we seek the hyperparameters 
\begin{equation}
\label{eqn:trainingobj}
 \!   \myVec{\theta}^{\star} =\! \mathop{\arg \max}\limits_{\myVec{\theta}} \sum_{\{\myMat{H}[b]\} \in \mySet{D}} \!R(\myMat{W}_J(\myVec{\theta}\textcolor{NewColor}{; \{\myMat{H}[b]\}})\myMat{A};  \{\myMat{H}[b]\}).
\end{equation}

The objective in \eqref{eqn:trainingobj} evaluates $\myVec{\theta}$  based on the output of the unfolded architecture, i.e., $\myMat{W}_J$. To facilitate the learning of $\myVec{\theta}$, we exploit the interpretability of the architecture, and particularly the fact that each layer should output a gradually improved modular beamformer. Accordingly, we propose a loss measure for optimizing $\myVec{\theta}$ that accounts for the output of the intermediate iterations (layers) in a monotonically growing fashion. Specifically, we use a loss function with logarithmic contribution growth between iterations~\cite{samuel2019learning,alter2024rapid}, given by
\begin{align}
\label{eqn:lossFunction}
    \mathcal{L}_{\mathcal{D}}(\myVec{\theta}) = 
  -&\sum_{\{\myMat{H}[b]\} \in \mySet{D}} \sum_{j=1}^{J} \frac{\log(1+j)}{|\mySet{D}|}\notag \\
  &\qquad \times  R(\myMat{W}_j(\myVec{\theta}\textcolor{NewColor}{; \{\myMat{H}[b]\}})\myMat{A};  \{\myMat{H}[b]\}).
\end{align}
The resulting training algorithm using mini-batch stochastic gradient descent is stated as Algorithm~\ref{alg:L2O}.
The negative sign in the loss  is added because our objective is to maximize the rate, rather than minimize it.

  \begin{algorithm}
    \caption{Learning Hyperparameters}
    \label{alg:L2O} 
    \SetKwInOut{Initialization}{Init}
    \Initialization{Set $\myVec{\theta}$ as fixed hyperparameters \newline Fix learning rate $\eta>0$ and  epochs $i_{\max}$}
    \SetKwInOut{Input}{Input}  
    \Input{Training set  $\mathcal{D}$}  
    {
        \For{$i = 0, 1, \ldots, i_{\max}-1$}{%
                    Randomly divide  $\mathcal{D}$ into $Q$ batches $\{\mathcal{D}_q\}_{q=1}^Q$\;
                    \For{$q = 1, \ldots, Q$}{
                   Apply model with parameters $\myVec{\theta}$ to $\mathcal{D}_q$\;                     
                    Update  $\myVec{\theta}\leftarrow \myVec{\theta} - \eta\nabla_{\myVec{\theta}}\mathcal{L}_{\mathcal{D}_q}(\myVec{\theta})$\;
                    }
                    
                    }
        \KwRet{$\myVec{\theta}$}
  }
\end{algorithm}
\color{NewColor}
\subsubsection{CSI Estimation Error Robustness}
\label{subssec:CSIRobust}
The iterative optimizer presented in this work assumes perfect \ac{csi}. However, in practice one may only have access to a noisy estimate of the \ac{csi}.
The fact that our unfolded \ac{pga} is trained from data allows it to also to learn hyperparameters that boost robustness to such noisy \ac{csi}, without altering the operation of the optimization algorithm.

The presence of \ac{csi} errors leads to having the optimizer applied to perturbed channel realizations, represented as ${\myMat{H}[b] + \myMat{E}[b]}$, rather than the true channel realizations ${\myMat{H}[b]}$. Here, $\myMat{E}[b]$ denotes the CSI estimation error. The trainability of unfolded \ac{pga} allows it to be trained with noisy \ac{csi} to yield beampatterns that are suitable for the true \ac{csi}. To that aim, we model the \ac{csi} error as an i.i.d. Gaussian noise matrix with zero mean and variance $\sigma_E^2$, which represents the expected level of estimation error. During training, where one has access to the channel realizations as in \ref{eqn:DataSet},  instead of applying the trainable optimizer to the true channel realizations, we apply
it to a noisy version of the channels.
To compute the loss during training,  the noiseless channels ${\myMat{H}[b]}$ are used to calculate the rate. This approach ensures that the optimizer learns hyperparameters that yield beamformers tailored to the true underlying channels, rather than to their noisy estimates. 

Mathematically, the loss function in \eqref{eqn:lossFunction} is updated as follows:
\begin{align}
\mathcal{L}_{\mathcal{D}}^{\sigma_E^2}(\myVec{\theta}) =&  
-\sum_{\{\myMat{H}_i[b]\} \in \mySet{D}} \sum_{j=1}^{J} \frac{\log(1+j)}{|\mySet{D}|} \notag \\
& \times
R(\myMat{W}_j(\myVec{\theta}; \{\myMat{H}_i[b]+\myMat{E}_i[b]\})\myMat{A};  \{\myMat{H}_i[b]\}), 
\label{eqn:lossFunctionCSI}
\end{align}
where $ \myMat{E}_i[b]\stackrel{iid}{\sim}\mathcal{N}(0,\sigma_E^2)$ is sampled anew for each forward pass during training. When $\sigma_E^2 = 0$, the training procedure reduces to the noiseless CSI case detailed in Subsection~\ref{subssec: Training}. The resulting noisy \ac{csi} aware training allows achieving improved robustness via hyperparameter learning, as empirically demonstrated in Subsection~\ref{subssec:resultsCSI}.  
%
\color{black}
\subsubsection{Transferability Across Configurations}
\label{ssec:Transfer}

Modular beamforming architectures are scalable,  stemming from the ability to achieve extremely large \ac{mimo} configurations by adding panels. A core gain of our interpretable learning-aided optimizers is the fact that they support such scalability. Specifically,  what we learn are hyperparameters of an optimization algorithm, which can be applied to different configurations. This indicates that our unfolded optimizer shares this transferability across configurations, thus inherently supporting large and adaptive scale deployments.

In particular, the architecture and training procedure detailed above are formulated for a given modular configuration. Nonetheless, a learned optimizer is transferable, and a learned optimizer trained for a small configuration can be scaled to applied to much larger ones. Specific scalable parameters of interest are the numbers of users $K$ and of panels $P$:
\begin{itemize}
    \item {\em Scaling $K$}: we note that the learned hyperparameters in \eqref{eqn:trainableParams} are invariant of the number of users. Thus, a learned optimizer trained for a small number of users can be readily applied to a setting with a large number of users.
    \item {\em Scaling $P$}: While the setting of the hyperparameters in \eqref{eqn:trainableParams} depends on the number of hyperparameters, as we allow panel-specific hyperparameters, one can scale by reusing hyperparameters across panels~\cite{noah2023limited}. For instance, a learned optimizer trained to set $\{\myMat{\alpha}_j, \myMat{\beta}_j\}$ for a configuration with $P$ panels can be applied to a setting with $c\cdot P$ a positive integer $c$, by e.g., setting its hyperparameters to be $\{\myMat{I}_c \otimes \myMat{\alpha}_j, \myMat{I}_c \otimes \myMat{\beta}_j \}$, where $\otimes$ denotes the Kronecker product, or by assigning the added panels an average of the learned hyperparameters.
\end{itemize}
The above approaches to transfer a learned optimizer across different configurations are empirically demonstrated to yield effective and scalable optimization in Subsection~\ref{ssec:results transfer}.

\subsection{Discussion}
\label{ssec:discussion}
The proposed unfolded \ac{pga}  is designed to learn from data how to traverse the optimization loss surface, allowing to achieve improved sum-rate within $J$ iterations, as shown in Section~\ref{sec:Experimental Study}.
The resulting learned $\myVec{\theta}$, which represents the per-iteration hyperparameters used for setting the modular beamformer, may exhibit certain counter-intuitive characteristics, such as negative gradient steps, as means of coping with non-convexity. As the resulting algorithm is a learned implementation of a principled optimizer, we share its scalability, which is highly desirable in modular architectures. 
While our formulation considers panels comprised phase-shifters, one can consider alternative hardware, such  as distributed active metasurfaces~\cite{shlezinger2019dynamic} and phase-time arrays~\cite{ratnam2022joint}. 

\textcolor{NewColor}{Our formulation of the modular hybrid beamforming setting considers the connectivity matrix $\myMat{A}$ to be fixed, representing a given physical infrastructure or wiring. Still, there are scenarios where it is of interest to also optimize $\myMat{A}$. These including the possibility of optimizing it for each channel alongside $\myMat{W}$, e.g., when the hardware includes rapidly adaptable switches, or even for having a single $\myMat{A}$ optimized for multiple coherence duration, thus effectively guiding the system design. While both options can be studied based on our proposed learned optimization methodology, we leave these extensions of our work for future study.}

The fixed and limited number of iterations impacts the computational complexity of setting the modular beamformers for a given channel realization. To quantify this complexity, we examine a single iteration of Algorithm \ref{alg:pga with momentum}. The computational burden is mainly dominated by the \textit{gradient computation} step, formulated in \eqref{eqn:dR_dW}. This computation has a complexity order (in complex multiplications) of
\begin{equation}
\mathcal{O}\Big(B(T^3 + M^2(T + K)) + M (T^2 + LPT + TK)\Big).
\end{equation}
The remaining steps in the algorithm include the \textit{update step} and the \textit{projection step} \eqref{eqn:Projection}. The complexity order of these steps is  $\mathcal{O}(MLP)$ for both, which is negligible compared to the gradient computation step. Considering that optimization procedure  operates over $B$ frequency bins, the overall complexity of the PGA+M algorithm with a predefined $J$ iterations, as implemented by the proposed learned optimizer, is of the order of
\begin{align}
\mathcal{C}_{PGA+M} = \mathcal{O}\Bigg(  JB\Big(&T^3 + M^2(T + K) 
 \notag\\
& + M (T^2 + LPT + TK)\Big)\Bigg).
\label{eqn:complexity}
\end{align}

The asymptotic complexity characterization in \eqref{eqn:complexity} is comprised of multiple summands, where in general none of the terms can be consistently ignored. Nonetheless, in an expected massive modular \ac{mimo} setting, where the overall number of antenna elements is larger than both the number of users ($K$) and the number of \ac{cpu} inputs ($T$), the dominating term is the one growing quadratically with $M$, and thus  $\mathcal{C}_{PGA+M} \approx \mathcal{O}(JBTM^2)$. 
When delving further into the analysis, it becomes evident that the dominant terms (in complex products) arise from matrix multiplication operations.
However, these computations can often be parallelized to reduce computational complexity, making the complexity of matrix inversion, represented by the $\mathcal{O}(JBT^3)$ term, the dominant factor. While it is possible to further reduce the number of multiplication operations through techniques such as sparse and low-rank approximations of the gradients, we leave this aspect to future research.

\section{Power-Aware Unfolded Modular Beamforming}
\label{sec:Power Aware Beamforming}

Our analysis in Section~\ref{sec:Unfolded Beamforming} focused on the unconstrained setting~\ref{itm:unconstratined}.
In this section, we use the unconstrained design for maximizing the rate \eqref{eqn:Rate} as a starting point, on top of which we develop a power-aware design. Specifically, we leverage the interpretability of the solution described in Section~\ref{sec:Unfolded Beamforming} to incorporate constraints \ref{itm:sparse}-\ref{itm:quant}. We focus on conventional (non-learned) gradient-based optimization subject to  \ref{itm:sparse}-\ref{itm:quant} in Subsections~\ref{ssec:Sparsified Modular Beamformer} and \ref{ssec:Quantized Modular Beamformer}, respectively. Then, we  formulate  \ac{pga} with momentum and the unfolded algorithms tailored for these designs in Subsections~\ref{ssec:PowerAwarePGA} and \ref{ssec:UnfoldedPowerAware}, respectively, and conclude with a discussion in Subsection~\ref{ssec:PowerAwareDiscussion}.

\subsection{Sparsified Modular Beamformer}
\label{ssec:Sparsified Modular Beamformer}

As detailed in Section~\ref{sec:System Model}, power efficiency can be enhanced by turning off some of the phase shifters.
Supporting deactivated phase shifters can be mathematically accommodated by constraint \ref{itm:sparse}, i.e., as promoting  sparsity of the matrix $\myMat{W}$. 
To meet this constraint, we rewrite our constrained optimization from \eqref{maximization_R} as 
\begin{align}
   \label{eqn: Sparse maximization_R}
    \myMat{W}^\star &= \mathop{\arg \max}\limits_{\myMat{W}= {\rm blkdiag}\{\myMat{W}_1,\ldots,\myMat{W}_P\}} R(\myMat{W}\myMat{A};  \{\myMat{H}[b]\}) \\
    &\text{subject to } ||\myMat{W}||_0 \leq S_{\max} \notag \\
    &\qquad\qquad \quad \myMat{W}_p \in \mySet{W}_{\rm sph}, \quad \forall p\in\{1,\ldots,P\}, \notag
\end{align}
where $S_{\max}  \leq N\cdot L \cdot P$ represents  the maximal allowed number of active phase shifters. 
In \eqref{eqn: Sparse maximization_R}, each phase shifter can be either active or turned off (by \eqref{eqn:PhasShiftBinary}). The $\ell_0$ constraint guarantees that at most $S_{\max}$ phase shifters are active.

Due to the inherent challenges associated with gradient-based optimization under $\ell_0$ constraints, we adopt the common approach of relaxing it into an $\ell_1$ constraint (see, e.g.,~\cite[Ch. 1]{eldar2012compressed}). The relaxed optimization problem is 
\begin{align}
   \label{eqn: Sparse maximization_R2}
    \myMat{W}^\star &= \mathop{\arg \max}\limits_{\myMat{W}= {\rm blkdiag}\{\myMat{W}_1,\ldots,\myMat{W}_P\}} R(\myMat{W}\myMat{A};  \{\myMat{H}[b]\})  - \lambda ||\myMat{W}||_1 \\
    &\text{subject to }   \myMat{W}_p \in \mySet{W}_{\rm sph}, \quad \forall p\in\{1,\ldots,P\}. \notag
\end{align}
In \eqref{eqn: Sparse maximization_R2},  $\lambda>0$ is an objective regularization coefficient~\cite{shlezinger2022model} whose purpose is to encourage the solution to hold the sparsity constraint in \eqref{eqn: Sparse maximization_R}. Setting the exact value of such hyperparameters to meet a desired sparsity level $S_{\max}$ is typically elusive, and involves repeating numerical trials to identify the exact value that strikes a desired trade-off between performance (i.e., rate) and sparsity. Nonetheless, once $\lambda$ is fixed, the gradient of the relaxed objective in \eqref{eqn: Sparse maximization_R2} can be computed as 
\begin{align}
    \nabla_\myMat{W} \Big(R(\myMat{W}\myMat{A};&  \{\myMat{H}[b]\}) -  \lambda||\myMat{W}||_1\Big) =\notag \\  &\nabla_\myMat{W} R(\myMat{W}\myMat{A}; \{\myMat{H}[b]\}) - \lambda \cdot e^{j\measuredangle\myMat{W}}.
    \label{eqn: grad with l1 norm}
\end{align}
where $\nabla_\myMat{W} R(\cdot)$ is given in Theorem~\ref{thm:dR_dW}.

Relaxing \eqref{eqn: Sparse maximization_R} into \eqref{eqn: Sparse maximization_R2} enables computing the objective gradients via  \eqref{eqn: grad with l1 norm}. In order to enable \ac{pga}-based optimization, one should also define the corresponding projection on the feasible set $\mathcal{W}_{\rm sph}$. To that end, we adopt a flexible extension of the projection operator in \eqref{eqn:Projection} to account for deactivated phase shifters, by introducing a magnitude threshold $\zeta>0$. The resulting projection step is given by 
\begin{equation}
\label{eqn:Sparse Projection}
[\mathcal{P}_{\mathcal{W}_{\rm sph}}(\myMat{W})]_{i,j} \!=
\! \begin{cases}
\frac{[\myMat{W}]_{i,j}}{\left|[\myMat{W}]_{i,j}\right|} \cdot
\delta_{\left\lfloor\frac{i}{N}\right\rfloor, \left\lfloor\frac{j}{L}\right\rfloor}
& |[\myMat{W}]_{i,j}| \geq \zeta, \\
0 & { |[\myMat{W}]_{i,j}| < \zeta}.
\end{cases}
\end{equation}
%
Projection via \eqref{eqn:Sparse Projection} is a crucial  step in the tackling the relaxation of \eqref{eqn: Sparse maximization_R}, ensuring that we set some elements of $\myMat{W}$ to $0$ and constrain their magnitude to be either $0$ or $1$, as described in \ref{itm:sparse}.
At the same time, this step preserves the block diagonal structure of $\myMat{W}$. A candidate choice of $\zeta$ is $\zeta = 0.5$; Setting $\zeta = 0$ makes this projection step coincide with the projection step in \eqref{eqn:Projection}.

\subsection{Quantized Modular Beamformer}
\label{ssec:Quantized Modular Beamformer}
A complementary approach to mitigate power consumption of analog modular processing is by discretizing the phase resolution with a few bit representation. Incorporating the resulting constraint \ref{itm:quant} into the rate maximization objective \eqref{maximization_R}, yields the resulting constrained optimization formulation
\begin{align}
   \label{eqn: Quant maximization_R}
    \myMat{W}^\star &= \mathop{\arg \max}\limits_{\myMat{W}= {\rm blkdiag}\{\myMat{W}_1,\ldots,\myMat{W}_P\}} R(\myMat{W}\myMat{A};  \{\myMat{H}[b]\}) \\
    &\text{subject to } \myMat{W}_p \in \mySet{W}^{Q}_{\rm qps}, \quad \forall p\in\{1,\ldots,P\}. \notag
\end{align}
In \eqref{eqn: Quant maximization_R}, the set of feasible phase shifts is restricted to take $Q$ quantization levels, using the definition of $\mySet{W}^Q_{\rm qps}$ in \eqref{eqn:PhasShiftQuant}. 

We note that the objective in \eqref{eqn: Quant maximization_R} is the same as in the unconstrained case \eqref{maximization_R}, and thus its gradient can be obtained from Theorem~\ref{thm:dR_dW}. Consequently, to formulate the suitable \ac{pga} steps for tackling \eqref{eqn: Quant maximization_R}, one must specify the quantization-aware projection operator. A direct approach to guarantee discrete phase components is by setting each phase element into its nearest discrete value, resulting in 
\begin{align}
\label{eqn:Quant Projection}
[\mathcal{P}_{\mathcal{W}_{\rm qps}^Q}(\myMat{W})]_{i,j} &= {e}^{j \phi_{i,j}^\star} \cdot \delta_{\left\lfloor\frac{i}{N}\right\rfloor, \left\lfloor\frac{j}{L}\right\rfloor}
\\ \notag
\text{where }\phi_{i,j} ^ \star &= \mathop{\arg \min}\limits_{\frac{\phi}{2\pi} \cdot Q \in \mathbb{Z}} 
|\phi - \measuredangle[\myMat{W}]_{i,j}|.
\end{align}
The projection operator in \eqref{eqn:Quant Projection} can be viewed as uniform scalar quantization of the phase of each element of $\myMat{W}$. 

\subsection{Power Aware PGA}
\label{ssec:PowerAwarePGA}

Here we present a baseline \ac{pga} algorithm for the updated set of constraints, as described in \ref{itm:sparse} and \ref{itm:quant}. 
This \ac{pga} algorithm is an extension of Algorithm \ref{alg:pga with momentum}, adapted to the power aware objectives described in Subsections \ref{ssec:Sparsified Modular Beamformer} and \ref{ssec:Quantized Modular Beamformer}.
The resulting formulation is summarized
as Algorithm \ref{alg:Power Aware pga with momentum}. There, we integrate the power-aware gradient formulation \eqref{eqn: grad with l1 norm} with the projection steps \eqref{eqn:Sparse Projection} and \eqref{eqn:Quant Projection} to realize \ac{pga}.

\begin{algorithm}
    \caption{Power-Aware PGA with Momentum}
    \label{alg:Power Aware pga with momentum} 
    \SetKwInOut{Initialization}{Init}
    \Initialization{Initialize $\myMat{W}_{-1}$ as zeros; randomize $\myMat{W}_{0}$;
    \newline Choose projection operator $\mySet{P}_{\mySet{W}}(\cdot)$;
    \newline Set iterations $J$, hyperparameters $\{\mu_j,\beta_j \}_{j=0}^{J-1} , \lambda$;.}
    \SetKwInOut{Input}{Input}  
    \Input{$\{\myMat{H}[b]\}_{b=1}^B$, $\myMat{A}$}
    {
        \For{$j = 0, 1, \ldots, J-1$}{%
                Set gradient $\nabla_\myMat{W}\big(R(\myMat{W}_j\myMat{A}) - \lambda ||\myMat{W}||_1\big)$ via \eqref{eqn: grad with l1 norm}\;
                Take gradient step $\myMat{W}_{j+1}\leftarrow \myMat{W}_j + \mu_j\nabla_\myMat{W}\big(R(\myMat{W}_j\myMat{A}) - \lambda ||\myMat{W}||_1\big)$\;
                %
                Add momentum  $\myMat{W}_{j+1}\leftarrow \myMat{W}_{j+1}+ \beta_j (\myMat{W}_j-\myMat{W}_{j-1})$\;
                Project $\myMat{W}_{j+1} \leftarrow \mySet{P}_{\mySet{W}}(\myMat{W}_{j+1})$  via \eqref{eqn:Sparse Projection} or \eqref{eqn:Quant Projection}\;  
                }
        \KwRet{$\myMat{W}_J$}
  }
\end{algorithm}

Algorithm ~\ref{alg:Power Aware pga with momentum} is designed to tackle the maximization problems outlined in \eqref{eqn: Sparse maximization_R2} and \eqref{eqn: Quant maximization_R} by proper settings of the projection operator and the regularization coefficient. For instance, by setting the projection to be given in \eqref{eqn:Quant Projection} and setting $\lambda = 0$, Algorithm~\ref{alg:Power Aware pga with momentum} tackles problem \eqref{eqn: Quant maximization_R}. Similarly, by using $\lambda > 0$ and setting the projection to be \eqref{eqn:Sparse Projection}, it tackles problem \eqref{eqn: Sparse maximization_R2}.


\subsection{Unfolded Power Aware PGA}
\label{ssec:UnfoldedPowerAware}
In Subsection \ref{ssec:unfoldedPGA}, we showed how data can be leveraged to tune the hyperparameters of Algorithm \ref{alg:pga with momentum} to enhance its performance within $J$ iterations, by converting it into a machine learning architecture.
Here, we follow this methodology to convert the power-aware Algorithm~\ref{alg:Power Aware pga with momentum} into a trainable architecture. As opposed to the unconstrained case, where the learned hyperparameters were solely associated with the solver, i.e., with \ac{pga}, in the following we also leverage data to tune parameters associated with the optimization objective~\cite{shlezinger2022model}, and particularly the (typically elusive) regularization coefficient $\lambda$.

\subsubsection{Architecture}
Following the conversion of \ac{pga} with momentum into a machine learning model  in Subsection~\ref{subssec:Architecture}, we view the iterative algorithm as a multi-layer architecture. This allows using the  data \eqref{eqn:DataSet} to learn  hyperparameters that yield desirable settings within a few iterations. 

Algorithm \ref{alg:Power Aware pga with momentum}  introduces another hyperparameter, $\lambda$, associated with optimization objective, and particularly the $\ell_1$ norm reduction. Here, we increase the abstractness of the trainable architecture by allowing $\lambda$ to differ from one iteration to another, i.e., have each iteration taking its gradient step over a {\em different loss surface}. Furthermore, we turn the scalar regularization coefficient into a matrix with different values for each entry in $\myMat{W}$. We add these learnable matrices $\{\myMat{\lambda}_j\}_{j=0}^{J-1}$ to the set of trainable parameters $\myVec{\theta}$. Accordingly, we now have  $3 \cdot J \cdot P \cdot N \cdot L$ trainable parameters, written as
\begin{equation}
\label{eqn:PA trainableParams}
    \myVec{\theta}=  \{\myMat{\alpha}_j, \myMat{\beta}_j, \myMat{\lambda}_j\}_{j=0}^{J-1}.
\end{equation}

\begin{figure}
    \centering
    \includegraphics[width=\columnwidth]{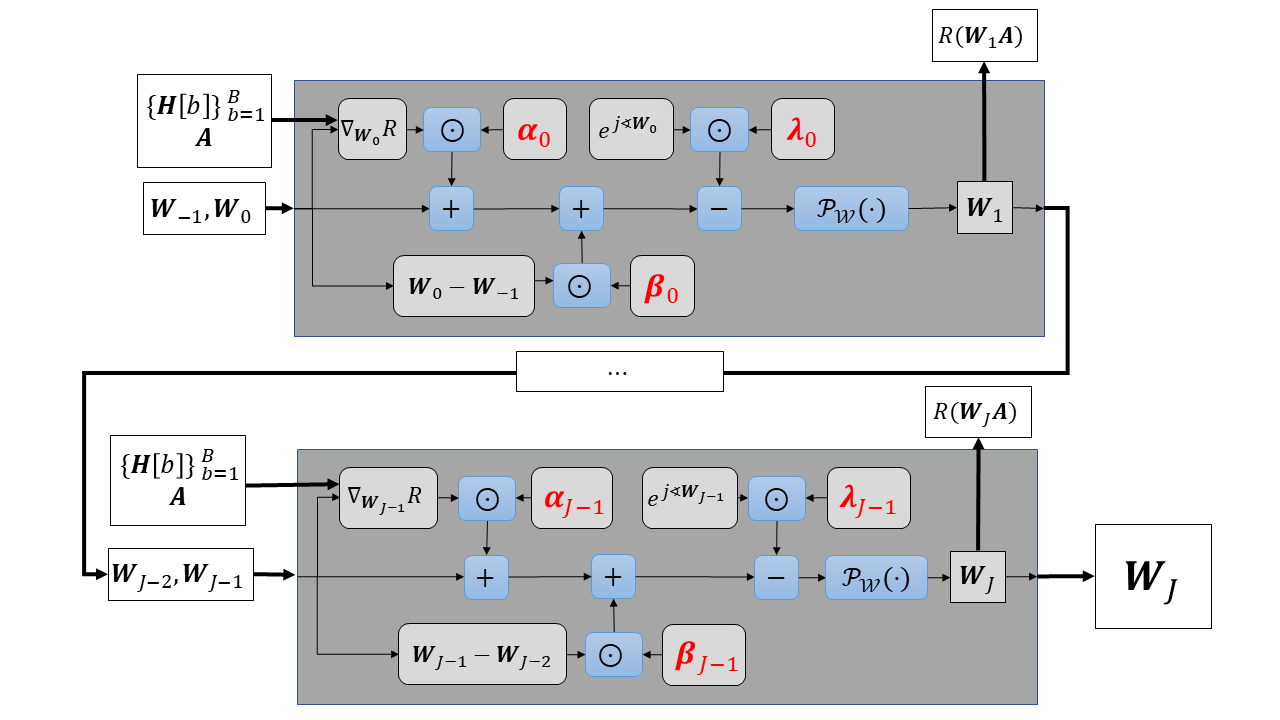}
    \caption{Power aware unfolded architecture, trainable parameters marked with \textcolor{red}{red}.}
    \label{fig:Power Aware Deep Unfolding}
\end{figure}

The power aware architecture is visualized in Fig.~\ref{fig:Power Aware Deep Unfolding} for a given learned hyperparameters $\myVec{\theta}$. For clarity of presentation, in Fig.~\ref{fig:Power Aware Deep Unfolding} we separate the gradient step in Algorithm~\ref{alg:Power Aware pga with momentum} into two separate steps, each one dependent on a different hyperparameter.

\begin{figure*}
    \centering
      \begin{subfigure}{0.7\columnwidth}
    \includegraphics[width=\columnwidth]{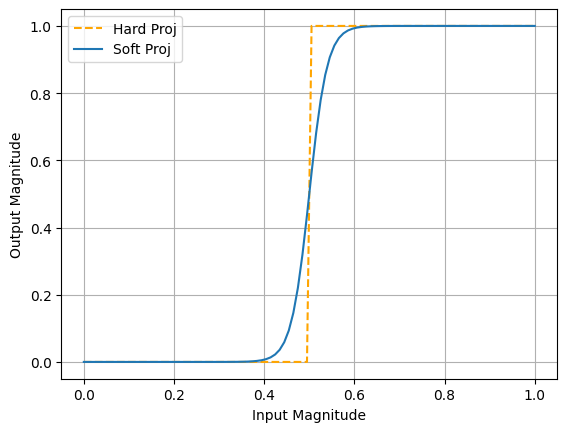}
    \caption{Magnitude projection, $\zeta = 0.5$.}
    \label{fig:Mag projection}
   \end{subfigure}
     \qquad      \qquad
   \begin{subfigure}{0.7\columnwidth}
    \includegraphics[width=\columnwidth]{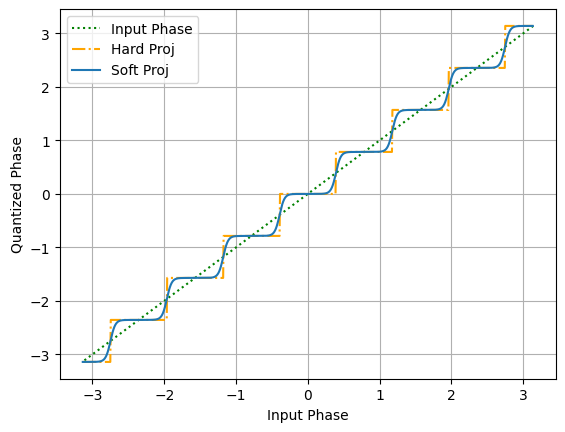}
    \caption{Phase projection, $Q=8$.}
    \label{fig:PhaseProjection}
    \end{subfigure}
    \vspace{0.4cm}
    \caption{Projection operators for power-aware \ac{pga} and their corresponding differentiable approximation.} 
\label{fig:projections}
\end{figure*}

\subsubsection{Projection}
A key issue in deep unfolding is maintaining the  forward path differentiable. This allows the optimizer to be represented as a  trainable architecture that can be tuned with deep learning techniques based on  backpropagation~\cite{shlezinger2022model}. Algorithm \ref{alg:Power Aware pga with momentum} mainly differs from Algorithm \ref{alg:pga with momentum} in the projection step. The projection step in both \eqref{eqn:Sparse Projection} and \eqref{eqn:Quant Projection} contains continuous-to-discrete mappings that nullify gradients.

To overcome this, we distinguish between the projection step utilized when applying the optimizer, and a  differentiable surrogate utilized for evaluating gradients during training. Specifically, we use the following surrogate projections:
\begin{itemize}
    \item {\em Approximating $\mathcal{P}_{\mathcal{W}_{\rm sph}}$}: The projection onto $\mathcal{W}_{\rm sph}$ can be viewed as a step function. Accordingly, we approximate it during training using a harp and shifted sigmoid function \cite{agustsson2017soft} denoted $\hat{\mathcal{P}}_{\mathcal{W}_{\rm sph}}$, given by
    \begin{equation}
        [\mathcal{P}_{\mathcal{W}_{\rm sph}}]_{i,j} = |\myMat{W}_{i,j}| = \sigma (s_m (|\myMat{W}_{i,j}| - \zeta_m)).
        \label{eqn:ApproxMagProj}
    \end{equation}
    In \eqref{eqn:ApproxMagProj}, $\sigma$ denotes the sigmoid function, $s_m$ denotes a sharpness hyperparameter, and $\zeta_m$ stands for the shift, imitating the threshold $\zeta$ from \eqref{eqn:Sparse Projection}.
    \item {\em Approximating $\mathcal{P}_{\mathcal{W}^Q_{\rm qph}}$}: For phase approximation with $Q$  quantization levels, we note that the projection operator is a piecewise-constant monotonically increasing function. Following~\cite{shlezinger2022deep}, we  approximate the quantization levels with sum of $Q$ differently shifted sigmoid functions. The resulting phase projection sets $\phi_{i,k}^{\star}$ in \eqref{eqn:Quant Projection} to be 
\begin{equation}
    \hat{\phi}_{i,k}^{\star}= \frac{2\pi}{Q} \sum_{q=1}^Q \sigma\Big(s_p\big(\measuredangle \myMat{W}_{i,j} - \frac{2\pi (q-0.5)}{Q} + \pi\big)\Big),
      \label{eqn:ApproxQProj}
\end{equation}
where $s_p$ is a sharpness hyperparameter.

\end{itemize}

This above formulations  approximate the  projection operators in a differentiable manner, ensuring smooth gradients during training while maintaining the characteristics of the original projection step.
 The ability of \eqref{eqn:Quant Projection}-\eqref{eqn:ApproxQProj}  to approximate their corresponding hard (non-differentiable) projection operators is illustrated in Fig.~\ref{fig:projections}.




\subsubsection{Loss Function}
We assume access to a dataset $\mySet{D}$, which comprises past channel realizations.
This data is leveraged to learn the hyperparameters for solving the following constrained optimization problem:
\begin{align}
\label{eqn:PowerAwareObj}
    &\myVec{\theta}^{\star} =  \mathop{\arg  \max}  \limits_{\myVec{\theta}}  \sum_{\{\myMat{H}[b]\} \in \mySet{D}} R(\myMat{W}_J(\myVec{\theta};  \textcolor{NewColor}{ \{\myMat{H}[b]\}})\myMat{A}; \{\myMat{H}[b]\}) \\ \notag
    &\text{subject to: }
    \myMat{W}_J= {\rm blkdiag}\{\myMat{W}_{J_1},\ldots,\myMat{W}_{J_P}\} \\ \notag
    &(i) \quad  ||\myMat{W}_J||_0 \leq S_{\max}, \quad  \myMat{W}_{J_p} \in \mySet{W}_{\rm sph}, \quad \forall p\in\{1,\ldots,P\}. \\ \notag
    &(ii) \quad \myMat{W}_{J_p} \in \mySet{W}_{\rm qps}, \quad \forall p\in\{1,\ldots,P\}. \notag
\end{align}
The constraints on $\myMat{W}$ are either  $(i)$ or $(ii)$, depending on the solved problem, \eqref{eqn: Sparse maximization_R} or in \eqref{eqn: Quant maximization_R}, respectively.

The learning objective in \eqref{eqn:PowerAwareObj} under $(i)$ encapsulates a key difference from  the unconstrained one  \eqref{eqn:trainingobj}.  Specifically, for sparse phase shifters, the learning objective stems from \eqref{eqn: Sparse maximization_R}, namely, maximizing $R$ under the $\ell_0$ norm constraint. When formulating the resulting \ac{pga} iterations, we relaxed $\ell_0$ norm to the $\ell_1$ norm, using the coefficient $\lambda$ to balance between rate maximization ($R$) and  sparsification. To translate this into a suitable loss for training that encourages a desired $\ell_1$ norm, we introduce the following training loss 
\begin{align}
    &\mathcal{L}_{\mathcal{D}}(\myVec{\theta}) = \notag \\ 
    &-\frac{1}{|\mySet{D}|}  \sum_{\{\myMat{H}[b]\} \in \mySet{D}}  \sum_{j=1}^{J} \log(1+j) R(\myMat{W}_j(\myVec{\theta}; \textcolor{NewColor}{\{\myMat{H}[b]\}})\myMat{A};  \{\myMat{H}[b]\}) \notag \\
     &- \gamma(\|\myMat{W}_J(\myVec{\theta};\textcolor{NewColor}{\{\myMat{H}[b]\}})\|_1 - S_{\max})^2,
    \label{eqn:TrainingLoss2}
\end{align}
where $\gamma \geq 0$ is a regularization hyperparameter. The loss \eqref{eqn:TrainingLoss2}  is also used with quantized phase shifters, i.e., under $(ii)$, by setting $\gamma=0$, as constraint $(ii)$ is guaranteed to hold when using the projection operator $\mathcal{P}_{\mathcal{W}^Q_{\rm qph}}$ in modular beamforming. 

The updated loss function \eqref{eqn:TrainingLoss2} is an extension of the unconstrained loss function \eqref{eqn:trainingobj}, adding the squared error between the $\ell_1$ norm of $\myMat{W}_J$ and $S_{\max}$. 
The rationale stems  from our desire to encourage $\|\myMat{W}_J\|_1$ (being a differentiable approximation of $\|\myMat{W}_J\|_0$) to approach $S_{\max}$ without having to tune different regularization coefficients for different $S_{\max}$. This allows training via, e.g., mini-batch stochastic gradient descent, as in Algorithm~\ref{alg:L2O}.


\subsection{Discussion}
\label{ssec:PowerAwareDiscussion}
The design of the constrained modular beamformer via Algorithm~\ref{alg:Power Aware pga with momentum} is based on the \ac{pga} steps designed for the unconstrained setting in Algorithm~\ref{alg:pga with momentum}, leveraging its interpretable operation by introducing dedicated projections. Despite the similarity in inference, the procedure involved in learning the hyperparameters from data substantially differs from the unconstrained case, necessitating differentiable approximations of the projection operators during learning, altering gradient computation, and a dedicated training loss. This allows learning to rapidly set constrained modular beamformers with only a minor degradation in the rate, as shown in Section~\ref{sec:Experimental Study}.

Once the hyperparameters are learned through the unfolded learning procedure detailed in Subsection~\ref{ssec:UnfoldedPowerAware}, the setting of the modular beamformer in Algorithm~\ref{alg:Power Aware pga with momentum} follows the unconstrained \ac{pga} in Algorithm~\ref{alg:pga with momentum}. The key differences are the additional phase term in the gradient computations via \eqref{eqn: grad with l1 norm}, and the modification of the projection operators via \eqref{eqn:Sparse Projection} or \eqref{eqn:Quant Projection}. As none of these operations dominates the computational burden, the complexity of the power-aware design is similar to that of the unconstrained case, analyzed in Subsection~\ref{ssec:discussion}. 

Our power-aware design boosts power efficiency by supporting deactivated and low-resolution phase shifters. The translation of these properties into  power savings is highly dependent on the hardware. We opt for a parametric representation of the desired sparsity level and phase shifter resolution, allowing the unfolded optimizer to be useful for different implementations and different power levels by setting of the corresponding parameters, e.g., $S_{\max}$ and $Q$. While our learning procedure utilizes the projection surrogates visualized in Fig.~\ref{fig:projections}, one can consider alternative differentiable approximations of such continuous-to-discrete mappings, see, e.g.,~\cite{lang2024data}. Moreover, additional methods to boost power efficiency by, e.g., deactivating panels rather than phase shifters, can be obtained as extensions to our framework, which are left for future study.



\section{Experimental Study}
\label{sec:Experimental Study}
In this section, we numerically evaluate the proposed unfolded algorithm for rapid tuning of uplink modular beamforming\footnote{The source code and hyperparameters are available at \url{https://github.com/levyohad/Power-Aware-Deep-Unfolding-for-Beamforming}}. We describe the experimental setup in Subsection~\ref{ssec:Experimental Setup}, while our results for unconstrained and power-aware settings are reported in Subsections~\ref{ssec:UnconstResults}-\ref{ssec:SparseResults}, respectively.

\subsection{Experimental Setup}
\label{ssec:Experimental Setup}
We simulate a communication system, where the channel matrices $\{\myMat{H}[b]\}_{b=1}^B$ are obtained from a Rayleigh fading distribution, as well as physically compliant realizations obtained from the \ac{quadriga} model \cite{6758357}. We treat the number of iterations $J$ as a pre-defined requirement, and train our algorithm to minimize its loss within this number of iterations.
The \ac{snr} is defined as  $1/\sigma_w^2$.

We compare our unfolded optimizer ({\em U-PGA+M}) with the following benchmarks: 
\begin{itemize} 
    \item {\em PGA + M:} running \ac{pga} with momentum (Algorithm~\ref{alg:pga with momentum}) with fixed hyperparameters $\{\mu,\beta\}, \forall j =1 \ldots J$, chosen from a discrete grid, to maximize the average $R$ after $J=500$ iterations for the entire training set. 
    \item {\em Line Search:}   \ac{pga} where the hyperparameters $\{\mu_j,\beta_j\}_{j=0}^{J-1}$ are selected for each iteration through grid search to maximize the achievable rate $R$ at each step~\cite{boyd2004convex}. This method performs an exhaustive search to find the optimal step size for each iteration, independently for each data sample, at the cost of excessive additional processing latency. 
    %
    \item {\em Manifold Optimization (MO):} The optimizer proposed in \cite{yu2016alternating} for centralized architectures, adapted to support   modular beamformers. It suggests a relaxed version of the maximization problem \eqref{maximization_R}, taking gradient steps of $\myMat{W}$ toward the fully digital beamformer $\myMat{G}_{\rm opt}$, trying to minimize the error $||\myMat{W}\myMat{A} - \myMat{G}_{\rm opt}||^2$, and projecting the resulted matrix $\myMat{W}$ on the feasible set, using $\mySet{P}_\mySet{W}(\cdot)$. As in the {\em Line Search} benchmark, it performs an exhaustive line search to choose the best step size for each iteration. 
   
    \item {\em \ac{cnn}:} A \ac{cnn} based on the architecture of \cite{dong2020framework}, trained to map channel realizations $\{\myMat{H}[b]\}_{b=1}^B$ into the phases of $\myMat{W}$ that maximize the channel-rate $R$. This approach is extremely parameterized. 
    For example, in a system with $K=7$ users, $M=8$ antennas, and $B=4$ frequency bins, the  \ac{cnn}  consists of $274088$ learned parameters, whereas {\em U-PGA+M} has only $960$ parameters.
\end{itemize}

\subsection{Unconstrained Modular Beamformers}
\label{ssec:UnconstResults}
We start the experimental study by considering unconstrained modular beamformers, as described in Subsection~\ref{ssec:unfoldedPGA}. Our evaluation first focuses on small scale \ac{mimo} settings, after which we assess transferability to larger systems, \textcolor{NewColor}{and then we examine the performance of our robust algorithm in tackling CSI uncertainty, demonstrating its ability to maintain reliability and efficiency under imperfect channel knowledge.}

\subsubsection{Small-Scale Hybrid Modular \ac{mimo}}
We first examine two scenarios representing relatively small-scale settings. In {\bf Scenario 1}, we set $(T=5, P=2, L=4, N=20, B=2, K=20)$, and the channel matrices obey Rayleigh fading. {\bf Scenario 2} uses \ac{quadriga} channels with $(T=5, P=4, L=2, N=3, B=4, K=5)$.  

\begin{figure}
    \centering
    \includegraphics[width=\columnwidth]{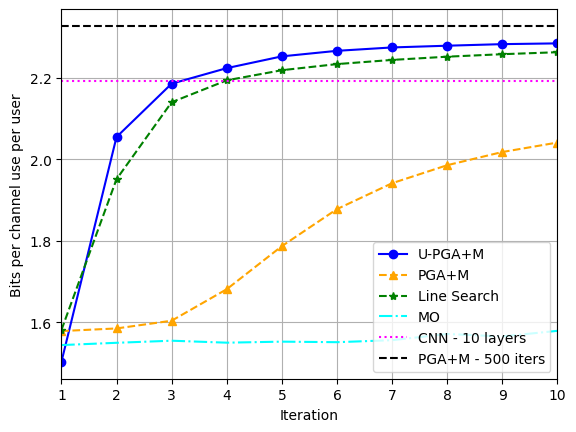}
    \caption{Sum-rate vs. iteration, {\bf Scenario 2}, ${\rm SNR}=0$ dB.}
    \label{fig:Results Scenario 2 - Quadriga unconstrained}
\end{figure}

We use these scenarios to evaluate the ability of our unfolded design to improve the sum-rate achieved within $J=10$ iterations. To that aim, we depict in Figs.~\ref{fig:Results Scenario 1 - i.i.d. synthetic channel, unconstrained}-\ref{fig:Results Scenario 2 - Quadriga unconstrained} the average sum-rate vs. the iteration index $j$ for {\bf Scenarios 1} and {\bf 2}, respectively.
\begin{figure}
    \centering
    \includegraphics[width=\columnwidth]{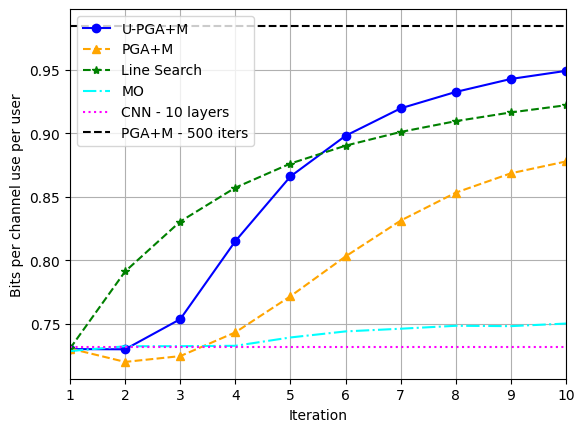}
    \caption{Sum-rate vs. iteration, {\bf Scenario 1}, ${\rm SNR}=0$ dB.}
    \label{fig:Results Scenario 1 - i.i.d. synthetic channel, unconstrained}
\end{figure}
For fair comparison, we set all optimization-based algorithms to have $J=10$ iterations, with the {\em \ac{cnn}} having $10$ layers, and compare the sum-rate to that achieved by {\em PGA+M}  with $J=500$ iterations. We consistently observe that in the considered scenarios with a limited number of iterations, our unfolded \ac{pga} outperforms all benchmarks, including the costly and greedy line search hyperparameter setting, and the highly parameterized \ac{cnn} benchmark. \textcolor{NewColor}{Specifically, the conventional {\em PGA+M} requires $6.7\times$ (for Rayleigh fading) and $14.7\times$ (for \ac{quadriga}) more iterations to achieve the performance of our unfolded algorithm.} The \ac{cnn} benchmark particularly struggles in Rayleigh channels, as \ac{cnn}s tends to capture spatial relationships, and such relationships do not exist under Rayleigh fading. Similarly, the MO optimizer exhibits slow convergence due to the additional modular structure imposed in our setting of modular beamformers. 

The results after $J$ iterations illustrated in Figs. \ref{fig:Results Scenario 1 - i.i.d. synthetic channel, unconstrained}-\ref{fig:Results Scenario 2 - Quadriga unconstrained} are also reported in Table~\ref{Tbl:RateVsSNR}, along with latency of the various methods. The latter is evaluated for all algorithms on the same platform (Google Colab premium). 
From Table~\ref{Tbl:RateVsSNR}, we observe that the unfolded algorithm, with $J = 10$ iterations, achieves a latency of $417 \mu s$. In contrast, optimizers employing line search require $123 \text{ ms}$, which is approximately $300 \times$ higher. Similarly, \ac{pga} running $J = 500$ iterations achieves comparable sum-rate performance but incurs a latency that is $50 \times$ greater. These results highlight the significant efficiency advantage of the proposed unfolded algorithm in achieving rapid convergence.
The {\em \ac{cnn}} benchmark has the least latency here, due to the efficient implementation and natural support of such architectures in platforms such as Google Colab.
This efficiency persists across multiple scenarios, emphasizing our ability to provide reliable modular beamformers with minimal latency, which can be further enhanced by employing dedicated hardware accelerators.
\begin{table}[]
\vspace{0.2cm}
\scriptsize
\centering
\begin{tabular}{cccccc}
\hline
\multirow{1}{*}{\bf Algorithm} & \multirow{1}{*}{\bf $J$} & \multicolumn{1}{c}{\bf Scenario 1} & \multicolumn{1}{c}{\bf Scenario 2} & \multicolumn{1}{c}{\bf Scenario 3} & \multirow{1}{*}{\bf Latency[$\mu s$]} \\ \hline
U-PGA+M          & 10    & 0.95     & 2.28    & 0.414   & 417    \\ \hline
\multirow{2}{*}{PGA+M}
                & 10     & 0.88     & 2.04    & 0.407   & 413     \\
                & 500    & 0.98     & 2.33    & 0.421   & 19352    \\ \hline
Line Search     & 10     & 0.92     & 2.26    & 0.415   & 123705   \\ \hline
CNN             & 10     & 0.73     & 2.19    & 0.388   & 290      \\ \hline
MO              & 10     & 0.75     & 1.58    & 0.388   & 31767    \\ \hline
\vspace{0.1cm}
\end{tabular}
\caption{Sum-rate and running time for different {\bf Scenarios 1}, {\bf 2}, and {\bf 3}. In {\bf Scenario 3},  \textit{U-PGA+M} was originally trained on {\bf Scenario 4}.
 \textit{Latency}  relates to the computation time of a single channel realization in \textbf{Scenario 2}. }
\label{Tbl:RateVsSNR}
\end{table}
\begin{figure}
    \centering
    \includegraphics[width=\columnwidth]{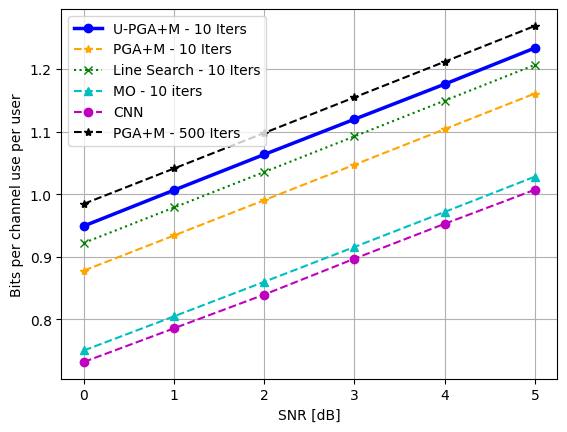}
    \caption{Sum-rate vs. SNR, {\bf Scenario 1}.}
    \label{fig:Scenario 1 - SNR}
\end{figure}
\begin{figure}
    \centering
    \includegraphics[width=\columnwidth]{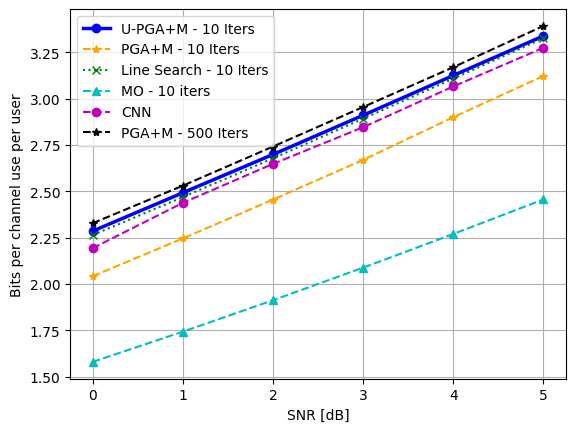}
    \caption{Sum-rate vs. SNR, {\bf Scenario 2}.}
    \label{fig:Scenario 2 - SNR}
\end{figure}
Figs. \ref{fig:Scenario 1 - SNR} and \ref{fig:Scenario 2 - SNR} show the resulted sum-rate for each benchmark in a wide range of \ac{snr} values, for {\bf Scenarios 1} and {\bf 2}, respectively. 
There, we observe that the gains obtained by the unfolded algorithm while using the same number of iterations, are translated into \ac{snr} gains of $1.25 - 1.5$ dB compared to the \textit{\ac{pga} + M} algorithm with the same $J$ iterations.
\subsubsection{Transferable Modular Structures Simulation Results}
\label{ssec:results transfer}
As explained in Subsection \ref{ssec:Transfer}, one of the key advantages of our learn-to-optimize modular beamforming design is the transferability of learned parameters when scaling the system. Modular structures allow us to train the system with specific settings and then scale it up to a larger number of users or panels without needing to retrain the model. 

\begin{figure}
    \centering
    \includegraphics[width=\columnwidth]{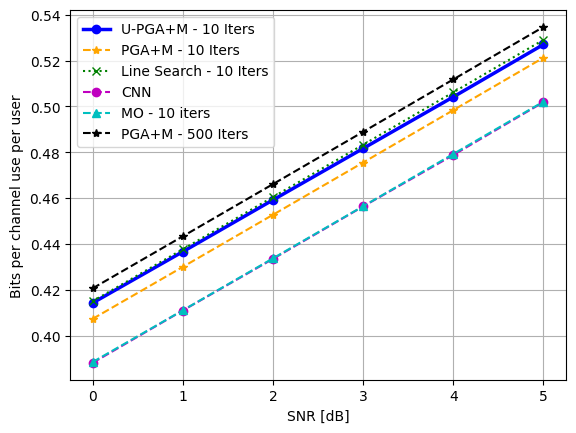}
    \caption{Sum-rate vs. SNR, {\bf Scenario 4} ($K=50$ users).  Here, {\it U-PGA+M} was trained on $K=5$ users. }
    \label{fig:Scenario 3 and 4 - SNR Scaling K}
\end{figure}

{\bf Scaling the number of users:}
To demonstrate this, we first evaluate {\em U-PGA+M} that was initially trained with a small number of users, $K=5$, and then deployed in a system with a large number of users, $K=50$, using the same hyperparameters.  This example illustrates the transferability of the unfolded model.
For this evaluation, we introduce {\bf Scenario 3}: $(T=5, P=8, N=4, B=2, K=50)$ and {\bf Scenario 4}: $(T=5, P=8, N=4, B=2, K=5)$. In both scenarios the channel realizations are sampled from a Rayleigh fading distribution.
Fig. \ref{fig:Scenario 3 and 4 - SNR Scaling K}, presents the sum-rates achieved by the various benchmarks  on {\bf Scenario 3}, along with the performance of the {\it U-PGA+M} algorithm, which was trained on {\bf Scenario 4}. These results demonstrate the flexibility of the unfolded algorithm and underscore its advantages in adapting to rapidly changing environments. The average rates at \ac{snr} of $0$  are also reported in Table~\ref{Tbl:RateVsSNR}.

{\bf Scaling the number of panels:}
Next, we provide an example of scaling up the number of panels. In this example, we use $c = 8$ as the scaling coefficient, and increase the number of panels from $P$ to $c\cdot P$, using $\{ \myMat{\alpha}_j, \myMat{\beta}_j \} = \{\myMat{I}_c \otimes \myMat{\alpha}_j, \myMat{I}_c \otimes \myMat{\beta}_j\}$.
To this end, we introduce {\bf Scenario 5}: $(T=5, P=2, N=4, B=2, K=7)$ and {\bf Scenario 6}: $(T=5 \cdot 8, P=2 \cdot 8, N=4, B=2, K=7)$; both  obey Rayleigh fading.
%
\begin{figure}
    \centering
    \includegraphics[width=\columnwidth]{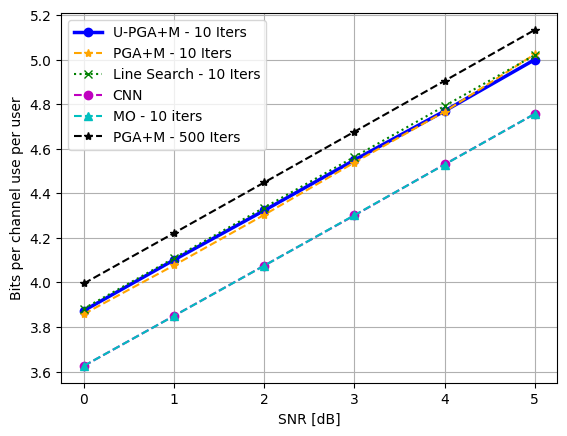}
    \caption{Sum-rate vs. SNR, {\bf Scenario 6}. 
    Here, {\it U-PGA+M} was trained on $M=8$, and deployed on $M=64$.}
    \label{fig:Scenario 5 and 6 - SNR Scaling P}
\end{figure}
Fig. \ref{fig:Scenario 5 and 6 - SNR Scaling P}
illustrates the results of various benchmarks trained on {\bf Scenario 6}, alongside the performance of the {\it U-PGA+M} algorithm that was trained on {\bf Scenario 5}. 
\textcolor{NewColor}{While the gains of our unfolded algorithm (trained for {\bf Scenario 5}) compared to the benchmarks that are all tuned for  {\bf Scenario 6} are more modest here compared with having it trained for the same setup, the fact that it does not degrade performance, and is even applicable in such different setups without necessitating any form of retraining, is quite non-trivial.}
These results highlight the flexibility of our design, demonstrating how it can be initially trained on a smaller network, perform straightforward calculations, and then be effectively scaled to much larger and more complex systems.
\color{NewColor}
\subsubsection{Robustness to CSI Estimation Error Simulation Results}
\label{subssec:resultsCSI}
As detailed in Subsection \ref{subssec:CSIRobust}, the trainabiltiy of our unfolded optimizer can be leveraged to cope with \ac{csi} estimation errors caused by e.g., noise or limited pilots. To evaluate the robustness of our unfolded beamforming algorithm, we simulate the impact of \ac{csi} estimation errors in {\bf Scenario 2} with $0$ dB \ac{snr}, and analyze its performance in terms of sum-rate as the variance of the estimation error increases.  

\begin{figure}
    \includegraphics[width=\columnwidth]{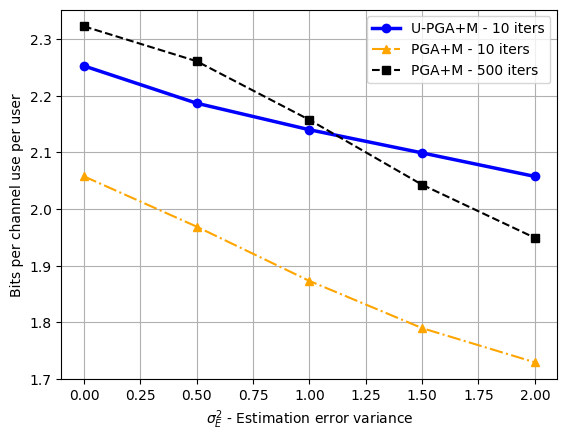}
    \caption{Sum-rate vs. CSI estimation error variance.}
    \label{fig:CSIError}
\end{figure}

The results, shown in Fig.~\ref{fig:CSIError}, highlight the comparative robustness of the proposed unfolded \ac{pga} algorithm (\textit{U-PGA+M}), which operates with solely 10 iterations. While performance degradation is observed for all methods as the estimation error variance (\(\sigma_E^2\)) increases, the unfolded \ac{pga} algorithm demonstrates a less steep descent compared to the conventional \textit{PGA+M} approach, even when the latter is executed with 500 iterations.
In particular,  \textit{PGA+M}  with 10 iterations exhibits the most significant performance degradation, indicating that insufficient optimization steps amplify the sensitivity to \ac{csi} estimation errors. In contrast, the \textit{PGA+M} method with 500 iterations reduces this sensitivity but still underperforms relative to the unfolded algorithm in scenarios with high \(\sigma_E^2\). This indicates the ability of our unfolded approach in achieving robustness imperfect \ac{csi} via dedicated learnin.  
The unfolded algorithm’s performance at lower \(\sigma_E^2\) remains competitive with conventional methods, confirming that the benefits of robustness do not come at the cost of baseline performance. Overall, these findings demonstrate that the proposed unfolded \ac{pga} framework not only provides robust performance under realistic channel conditions, but also offers a scalable solution for handling imperfect \ac{csi} in modular beamforming systems.

\color{black}
\subsection{Power-Aware Modular Beamformers}
\label{ssec:SparseResults}
We proceed to evaluating the ability of our unfolded optimizer to encourage power-efficient modular beamformers. We divide our evaluation to boosting efficiency via deactivating phase shifters, and via using low resolution phase shifters.

\subsubsection{Sparse Modular Beamformers}
We evaluate the effect of designing modular beamformers with deactivated phase shifters by boosting sparsity in $\myMat{W}$. We focus on {\bf Scenario 2},  and apply {\em U-PGA+M} using Algorithm \ref{alg:Power Aware pga with momentum}  utilizing the projection step  $\mySet{P}_\mySet{W}(\cdot)$ in  \eqref{eqn:Sparse Projection}. We set the parameter $S_{\max}$ to be $0.75 \cdot N \cdot L \cdot P$. 
As the benchmarks detailed in Subsection~\ref{ssec:Experimental Setup} do not naturally account for such constraints, we compare here {\em U-PGA+M} to unconstrained settings that are based on \ac{pga}, possibly with line search and/or large number of iterations.

\begin{figure}
    \centering
    \includegraphics[width=\columnwidth]{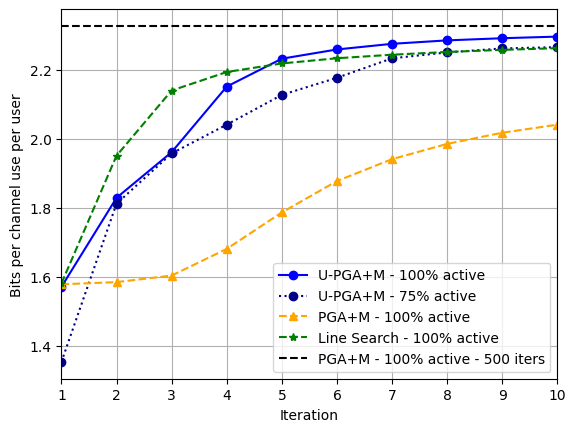}
    \caption{Sum-rate vs. iteration, {\bf Scenario 2}, ${\rm SNR}=0$ dB.}
    \label{fig:Results Scenario 2 - Quadriga Sparse}
    \end{figure}

\begin{figure}
    \centering
    \includegraphics[width=\columnwidth]{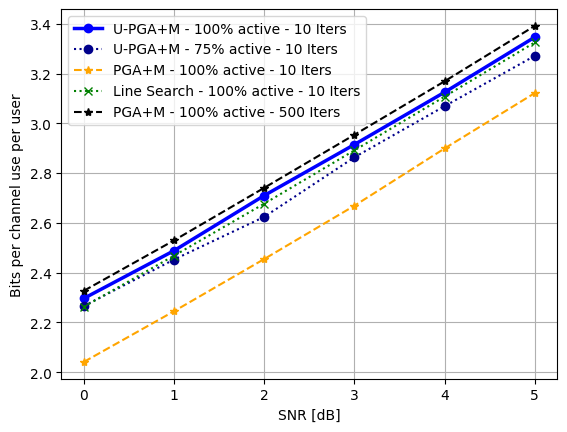}
    \caption{Sum-rate vs. SNR, {\bf Scenario 2}.}
    \label{fig:Scenario 2 - SNR Sparse}
\end{figure}

The resulting rates vs. iteration compared to those achieved without such power-aware design are reported in Fig. \ref{fig:Results Scenario 2 - Quadriga Sparse}. We observe that by applying Algorithm \ref{alg:Power Aware pga with momentum}, one achieves a $25 \%$ reduction in active components while maintaining a comparable sum-rate and identical running time to unconstrained settings. 
The unfolded algorithm learns to rapidly compensates for the inactive components by tuning the active phase shifters.
Fig. \ref{fig:Scenario 2 - SNR Sparse} further highlights the advantages of the unfolded power-aware algorithm by evaluating the resulting sum-rate over different \acp{snr}. There, we observe that our power-aware design not only allows accurate operation with $25\%$ of the phase shifters deactivated, but it also offers approximately $1$ dB improvement in \ac{snr} over the \textit{\ac{pga} + M} benchmark  with the same number of iterations.
%

\subsubsection{Low-Resolution Modular Beamformers}
We conclude our numerical study by evaluating the effect of using quantized phase shifters, and the ability of our power-aware design to account for such constraint when tuning modular beamformers. 
We consider the power-aware  Algorithm \ref{alg:Power Aware pga with momentum}, utilizing the projection step for $\mySet{P}_\mySet{W}(\cdot)$ as specified in \eqref {eqn:Quant Projection}, when the phase shifters can be configured to merely $4$ bits in phase, i.e.,  $Q=16$. 
The benchmarks detailed in Subsection~\ref{ssec:Experimental Setup} are adapted to produce low-resolution phase shifters by altering the projection step (for optimization methods) or quantizing the produced phases (for the \ac{cnn}).  

\begin{figure}
    \centering
    \includegraphics[width=\columnwidth]{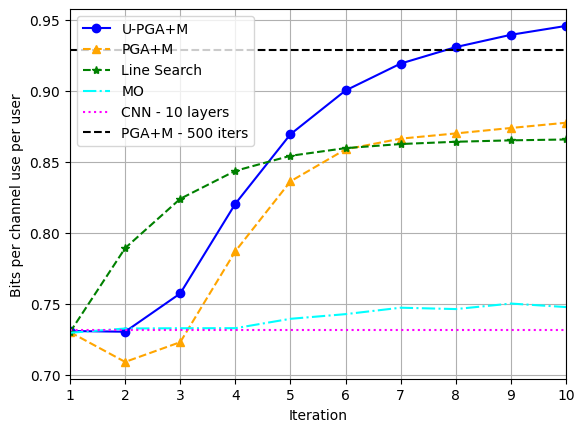}
    \caption{Sum-rate vs. iteration, {\bf Scenario 1}, 4-bit  phase shifters, ${\rm SNR}=0$ dB.}
    \label{fig:Results Scenario 1 - IID Quant}
    \end{figure}

\begin{figure}
    \centering
    \includegraphics[width=\columnwidth]{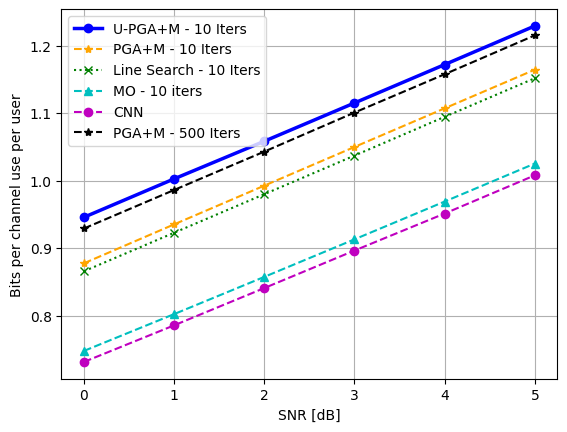}
    \caption{Sum-rate vs. SNR, {\bf Scenario 1}, 4-bit  phase shifters.}
    \label{fig:Scenario 1 - SNR IID Quant}
\end{figure}

The resulting sum-rate vs. iteration and vs. \ac{snr} for {\bf Scenario 1} are illustrated in Figs. \ref{fig:Results Scenario 1 - IID Quant} and \ref{fig:Scenario 1 - SNR IID Quant}, respectively.
The results demonstrate that the proposed unfolded algorithm, which is particularly geared to accounting for low-resolution phase shifters, significantly outperforms other benchmarks, including both slow iterative methods that are not trained as machine learning models, and suffer more notable degradation due to quantization, as well as the  heavily parameterized \ac{cnn}.
Specifically, while {\it \ac{pga} + M} struggles to overcome the quantization constraints using gradient steps, and tends to converge to a local maximum \textcolor{NewColor}{even when allowed to run for $500$ iterations}, the unfolded algorithm effectively learns to traverse the loss surface and projection constraints, achieving a better sum-rate. As shown in Fig. \ref{fig:Scenario 1 - SNR IID Quant}, the unfolded algorithm consistently performs better across various \ac{snr} values compared to all other benchmarks.

%


\section{Conclusions}
\label{ssec:conclusions}

In this work, we investigated rapid beamforming design for uplink modular \ac{mimo} systems under three different sets of constraints: unconstrained phase shifters, sparse phase shifters, and quantized phase shifters. Our proposed algorithm, a momentum-aided gradient-based optimization of the sum-rate objective, boosts power efficient settings while operating  with a fixed and small number of iterations. By transforming the optimizer into a lightweight, trainable model, we harness data to fine-tune its hyperparameters, enhancing its adaptability to various system conditions. Our experimental results demonstrate a significant improvement in the achieved sum-rate compared to benchmarks operating within the same computational-time constraints, while also surpassing other benchmarks in hardware-aware and power-aware designs.

\ifFullVersion
\begin{appendix}
\numberwithin{lemma}{subsection} 
\numberwithin{corollary}{subsection} 
\numberwithin{remark}{subsection} 
\numberwithin{equation}{subsection}	

\subsection{Proof of Theorem~\ref{thm:dR_dW}}
\label{app:Proof1}
For ease of presentation, we define the following auxiliary variables: 
$\myMat{G^\dag} \triangleq (\myMat{G}^H\myMat{G})^{-1}\myMat{G}^H$,
$\myMat{P} \triangleq \myMat{G} \myMat{G^\dag}$, 
$\myMat{Z}[b] \triangleq \myMat{P} \myMat{H}[b]\myMat{H}^H[b]$,
and $\myMat{Q}[b] \triangleq \myMat{I}_k + \frac{\rho_s}{\sigma_w^2} \myMat{Z}[b]$.
Using these symbols, the objective with respect to which the derivative is taken is written as
  $  R\left(\myMat{G}; \{\myMat{H}[b]\}\right) =  \frac{1}{B}\sum_{b=1}^B \log  \big|\myMat{Q}[b]\big|$.    
%
Substituting auxiliary variables, we aim at proving \eqref{eqn:dR_dW} by  showing that 
\begin{align} 
    &\nabla_\myMat{W} R(\myMat{W},\myMat{A}) = \notag \\ 
    &\qquad \frac{1}{B}\sum_{b=1}^B
    \bigg(\frac{\rho_s}{\sigma_w^2}\myMat{A}
    \myMat{G^\dag}\myMat{H}[b]\myMat{H}[b]^H
    \myMat{Q}^{-1}
    (\myMat{I}-\myMat{P})\bigg) ^H.
    \label{eqn:dR_dWL}
\end{align}
The representation in \eqref{eqn:dR_dWL} can be derived using the differential method for matrices as described in \cite{magnus99}.
In the following proof the sign $\textbf{:}$ denotes the Frobenius inner product, and the derivations are done using chain rule. For brevity we include our derivation for a single frequency bin.

Since $R =\log\big|\myMat{Q}\big|$ when $B=1$, it holds that the differential can be expressed as
\begin{align}
    dR =& \Big(\myMat{Q}^{-1}\Big)^H \; \textbf{:} \; d\myMat{Q} 
     \stackrel{(a)}{=} \Big(\frac{\rho_s}{\sigma_w^2}\myMat{Q}^{-1}\Big)^H \; \textbf{:}\; d\myMat{P} \, \myMat{H}\myMat{H}^H,
     \label{eqn:proof1}
\end{align}
where $(a)$ stems from $d \myMat{Q} = \cancelto{0}{d \myMat{I}_K} + d \frac{\rho_s}{\sigma_w^2} \myMat{Z} = \frac{\rho_s}{\sigma_w^2} d \myMat{Z}$, 
and from $d \myMat{Z} = d \myMat{P} \myMat{H}\myMat{H}^H,$ as $\myMat{H}$ is independent of $\myMat{W}$.

Next, we substitute  $\myMat{P}$, $\myMat{G}^\dag$, and $\myMat{G}$ using the chain rule:
\begin{align}
    dR = & \Big(\frac{\rho_s}{\sigma_w^2}\, (\myMat{H}\myMat{H}^H)\myMat{Q}^{-1}\Big)^H \; \textbf{:}\;(\myMat{I} - \myMat{P}) \, d\myMat{G} \: \myMat{G}^\dag.
    \label{eqn:derive_P}
\end{align}
Equation \eqref{eqn:derive_P} can be explained by recalling that $  \myMat{P} \triangleq \myMat{G} (\myMat{G}^H \myMat{G})^{-1} \myMat{G}^H$, and thus 
\begin{align*}
    d\myMat{P} &= d\myMat{G} (\myMat{G}^H \myMat{G})^{-1} \myMat{G}^H + \myMat{G} d(\myMat{G}^H \myMat{G})^{-1} \myMat{G}^H  \notag \\
    &\quad \quad \quad \quad + \myMat{G}(\myMat{G}^H \myMat{G})^{-1} \cancelto{0}{d\myMat{G}^H} \notag \\ \notag
     &= d\myMat{G} \, \myMat{G}^\dag - \myMat{G} (\myMat{G}^H \myMat{G})^{-1} \, \big(\cancelto{0}{d\myMat{G}^H}\myMat{G} + \myMat{G}^H d\myMat{G}\big)\, \myMat{G}^\dag  \\ \notag
 &= (\myMat{I} - \myMat{P}) \, d\myMat{G} \, \myMat{G}^\dag.
\end{align*}
Once \eqref{eqn:derive_P} is achieved, the rest of the proof becomes straight forward, using the chain rule until \eqref{eqn:dR_dWL} is reached, i.e., 
\begin{align*}
    dR =& \Big(\frac{\rho_s}{\sigma_w^2}\, \myMat{G}^\dag\myMat{H}\myMat{H}^H\myMat{Q}^{-1} (\myMat{I} - \myMat{P})\Big)^H \; \textbf{:}\; \, d\myMat{G}   \notag \\ 
    =&\Big(\frac{\rho_s}{\sigma_w^2}\, \myMat{A} \myMat{G}^\dag\myMat{H}\myMat{H}^H\myMat{Q}^{-1} (\myMat{I} - \myMat{P})\Big)^H \; \textbf{:}\; \, d\myMat{W} ,
\end{align*}
which concludes that $\nabla_\myMat{W} R(\myMat{W},\myMat{A}) = \big(\frac{\rho_s}{\sigma_w^2}\myMat{A}\myMat{G^\dag}\myMat{H}\myMat{H}^H\myMat{Q}^{-1}(\myMat{I}-\myMat{P})\big) ^H$, thus proving  \eqref{eqn:dR_dW}. 
\end{appendix}
\fi 
 
\bibliographystyle{IEEEtran}
\bibliography{IEEEabrv,refs}

\end{document}